\documentclass[useAMS,usenatbib,letterpaper]{mn2e}
\usepackage[totalwidth=480pt,totalheight=680pt,layoutvoffset=0.5cm]{geometry}
\usepackage{graphicx}
\usepackage{amsmath}
\usepackage{astrobib_mnras2e}
\usepackage{times}
\usepackage{fixltx2e}
\usepackage{lineno}

\title[Redshift weights for BAO : Methods and Mock Results]{Redshift Weights for Baryon Acoustic Oscillations : Application to Mock Galaxy Catalogs}

\author[Zhu et al]{Fangzhou Zhu$^1$, 
Nikhil Padmanabhan$^1$, 
Martin White$^2$,
Ashley J. Ross$^3$,
Gongbo Zhao$^{4,5}$ \\
\scriptsize $^1$ Dept. of Physics, Yale University, New Haven, CT 06511 \\
\scriptsize $^2$ Dept. of Physics and Astronomy, U.C. Berkeley, Berkeley, CA \\
\scriptsize $^3$ Center for Cosmology and Astroparticle Physics, Department of Physics, The Ohio State University, OH 43210, USA \\
\scriptsize $^4$ National Astronomy Observatories, Chinese Academy of Science, Beijing, 100012, P.R.China \\
\scriptsize $^5$ Institute of Cosmology \& Gravitation, University of Portsmouth, Dennis Sciama Building, Portsmouth, PO1 3FX, UK \\
}

\begin{document}

\maketitle
\begin{abstract}

Large redshift surveys capable of measuring the Baryon Acoustic
Oscillation (BAO) signal have proven to be an effective way of measuring
the distance-redshift relation in cosmology. Future BAO
surveys will probe very large volumes, covering wide ranges in
redshift.
Building off the work in \cite{Zhu15}, we develop a technique to directly constrain the distance-redshift relation
from BAO measurements without splitting the sample into redshift bins. 
We parameterize the distance-redshift relation, relative to a fiducial model, as a quadratic expansion. 
We measure its coefficients and reconstruct the distance-redshift relation from the expansion.

We apply the redshift weighting technique in \cite{Zhu15} to the clustering of galaxies 
from 1000 QuickPM (QPM) mock simulations after reconstruction and achieve a $0.75\%$
measurement of the angular diameter distance $D_{A}$ at $z=0.64$ 
and the same precision for Hubble parameter $H$ 
at $z=0.29$. These QPM mock catalogs are designed to mimic the clustering and noise level of the 
Baryon Oscillation Spectroscopic Survey (BOSS) Data Release 12 (DR12). 
We implement the redshift weights proposed in \cite{Zhu15} to compress the correlation functions 
in the redshift direction onto a set of weighted correlation functions. These estimators give unbiased $D_{A}$ and $H$ 
measurements at all redshifts within the range of the combined sample. 
We demonstrate the effectiveness of redshift weighting in improving the distance and Hubble parameter estimates. 
Instead of measuring at a single `effective' redshift as in traditional analyses, we report our $D_{A}$ and $H$ measurements at all redshifts. 
The measured fractional error of $D_{A}$ ranges from $1.53\%$ at $z=0.2$ to $0.75\%$ at $z=0.64$. 
The fractional error of $H$ ranges from $0.75\%$ at $z=0.29$ to $2.45\%$ at $z=0.7$. 
Our measurements are consistent with a Fisher forecast to within $10\%$ to $20\%$ depending on the pivot redshift. 
We further show the results are robust against the choice of fiducial cosmologies, galaxy bias models, and Redshift Space Distortions (RSD) streaming parameters. 
 
%We apply the redshift weighting described in \cite{Zhu15} to Baryon Acoustic Oscillation (BAO) measurements in the correlation function. 
%We develop and validate the method on the mock catalogs created for the Baryon Oscillation Spectroscopic Survey (BOSS) Data Release 12 
%combined sample covering $0.2 < z < 0.7$. 
%We demonstrate the efficiency of redshift weighting in improving the accuracy of 
%distance and Hubble parameter measurements. We examine the robustness of our analysis against the choice of
%fiducial cosmologies, pivot redshifts and galaxy bias models. 
\end{abstract}

\section{Introduction}

Baryon acoustic oscillations (BAO) are a geometrical probe of 
the universe via a standard ruler provided by the `baryon acoustic scale',
a characteristic scale imprinted in the distribution of galaxies \citep{Sunyaev&Zeldovich1970, Peebles&Yu1970, Bond&Efstathiou1987, Hu&Sugiyama1996, Eisenstein&Hu}.
Mapping the distribution of galaxies on large scales, 
one finds that galaxies are slightly more likely to be separated by a distance of roughly 150 Mpc. 
In the hot and ionized Universe at early times, 
photons and baryons are tightly coupled through Thomson scattering. 
The strong radiation pressure
pushes the photon-baryon fluid outwards in a spherical sound wave. 
Gravity, on the other hand, provides an inward
restoring force. This competition between matter and radiation 
gives rise to acoustic waves within the fluid. Once recombination happens, the
baryons and photons quickly decouple from each other. 
Photons quickly stream away from the baryons to form the
cosmic microwave background (CMB).
The acoustic waves then `freeze out' as the Universe becomes neutral as it expanded and cooled. 
Slight density enhancements at a scale set by the acoustic scale - 
distance an acoustic wave can travel between the time of the Big Bang and recombination - 
is magnified by gravitational interaction to seed the galaxy formation. The acoustic scale 
becomes a physical scale imprinted in the CMB and
is measurable in the clustering of galaxies today. 

Since its first detection \citep{Cole2005, Eisenstein05} a decade ago, 
BAO has been a prominent probe featured in a host of 
galaxy redshift surveys \citep{Blake2007, Kazin2010, Percival2010, Beutler2011, Padmanabhan2012, AndersonDR11}. Large surveys like BOSS \citep{Dawson13, Alam15}, 
a part of the Sloan Digital Sky Survey \citep{EisensteinDesign} 
have been pushing the measurement of the acoustic scale to
ever higher precision,
providing tighter constraints on our cosmological models. 

In current and future generations of BAO surveys, the samples cover
a wide range of redshift. 
In traditional analyses, one improves the resolution of the distance-redshift relation measurement by
splitting samples into multiple redshift bins and analyze the signals in these narrower slices. 
Such a splitting scheme has several disadvantages : (1) the signal-to-noise ratio is lower in each thin 
slice, (2) the choice of bins is often arbitrary, and (3) one loses signal across boundaries of disjoint bins.

To tackle the problems with binning outlined above, \cite{Zhu15} proposed
using a set of weights to compress the information in the redshift direction onto a 
small number of modes. These modes are designed to efficiently constrain the distance-redshift
relation parametrized in a simple generic form over the entire redshift extent
of the survey. This paper applies the methods proposed in \cite{Zhu15} to BOSS mock galaxy catalogs.
Our goal here is to demonstrate the practicability, robustness and efficiency of the method. 

The paper is structured as follows: \S \ref{sec:theory} introduces the redshift weights and 
covers the basics of correlation function multipoles. \S \ref{sec:simulations} describes the simulations used in this work. 
In \S \ref{sec:analysis}, we describe the redshift weighting algorithm 
in detail and provide the fitting model. We
discuss the improvement in the fitting of the BAO feature in \S \ref{sec:results}. We conclude
in section \S \ref{sec:discussion} with a discussion of our results.

\section{Theory}
\label{sec:theory}
\subsection{Distance Redshift Relation}
\label{sec:distance_redshift}

In BAO analyses, one typically assumes a fiducial cosmology to convert the galaxy angular positions 
and redshifts into 3D positions and parametrizes deviations from this fiducial cosmology.
We follow the parametrization proposed in \cite{Zhu15}. 
We denote the comoving radial distance by $\chi(z)$. Choosing a pivot redshift $z_0$ within
redshift range of the survey, we express the ratio of the true and fiducial radial comoving distance
$\chi(z)/\chi_f(z)$ as a Taylor series in $x(z) \equiv\chi_f(z)/\chi_f(z_0) - 1$,
\begin{equation}
\label{eq:chi_ratio}
	\frac{\chi\left(z\right)}{\chi_{f}\left(z\right)}=\alpha_{0}\left(1+\alpha_{1}x+ \frac{1}{2}\alpha_{2}x^{2}\right).
\end{equation}
When the fiducial cosmology matches
the true cosmology, one will measure $\alpha_0 = 1$, $\alpha_1 = 0$, and $\alpha_2 = 0$. 

We can very easily extend this Taylor series to higher orders, but
the order chosen here is sufficient for wide deviations in the distance-redshift relation \citep{Zhu15}. 
We will discuss selecting the appropriate number of parameters later in the paper. 
The ratio between the fiducial and true Hubble parameter $H = 1/\chi'(z)$ is given by
\begin{equation}
\label{eq:h_ratio}
\frac{H_f(z)}{H(z)} = \alpha_0 \left[1+\alpha_1 + (2\alpha_1 + \alpha_2) x + \frac{3}{2}\alpha_2 x^2 \right].
\end{equation}
The parameters $\alpha_0$, $\alpha_1$ and $\alpha_2$ can be related to the true distance-redshift relation as 
\begin{align}
	\alpha_0 &= \frac{\chi_0}{\chi_{f,0}}   \\
	\alpha_1 &= \frac{H_{f,0} \chi_{f,0}}{H_0 \chi_0} - 1   \\
	\alpha_2 &= (1+\alpha_1) \chi_{f,0} [H'_{f,0} - \alpha_0(1+\alpha_1) H'_0] - 2\alpha_1. \label{eq:a2_theory}
\end{align}
Measuring $\alpha_0$, $\alpha_1$ and $\alpha_2$ allows one to reconstruct the distance-redshift relation according to 
Eq. \ref{eq:chi_ratio}.

\begin{figure*}
\centering 
\includegraphics[width=\textwidth]{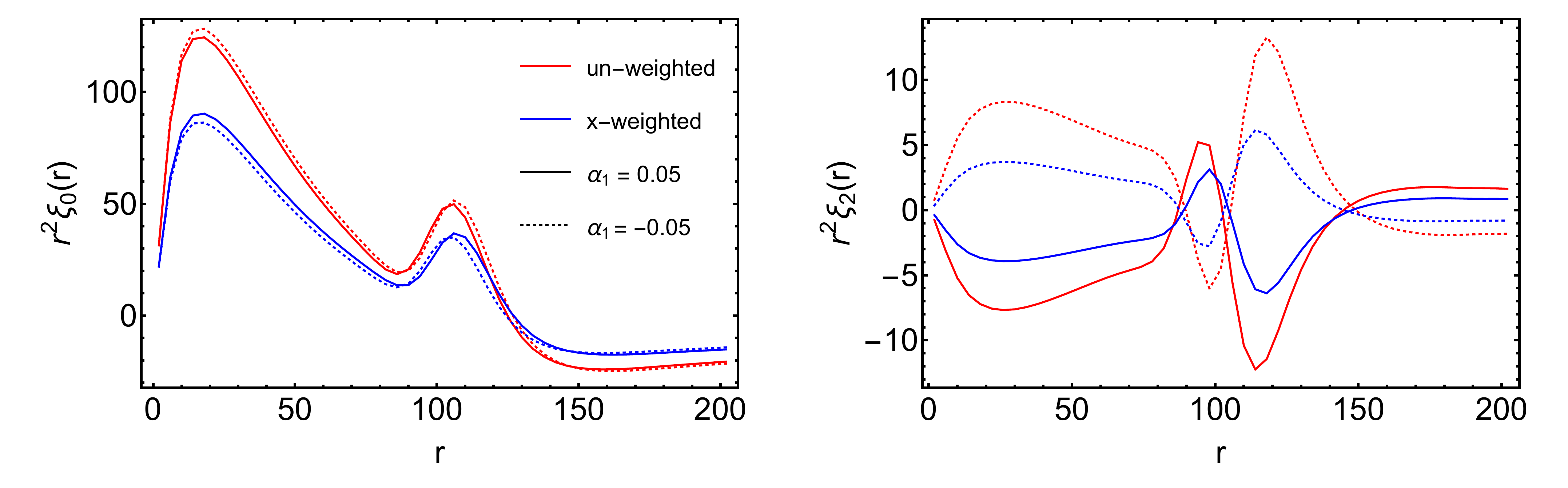}
\caption{The model monopole (left panel) and quadrupole (right panel) with a varying $\alpha_1$. 
The red lines plot the un-weighted monopoles and quadrupoles while the blue lines are for the $x$-weighted. 
The solid lines correspond to $\alpha = 0.05$ while the dotted lines are for $\alpha = -0.05$. 
For plotting convenience, we have multiplied the $x$-weighted monopoles and quadrupoles
by a factor of $-5$. 
All curves assume a pivot redshift of $z_0 = 0.57$ and fiducial model parameters 
$\Sigma_\perp = \Sigma_\parallel = 4.3 h^{-1}{\rm Mpc}$ with $\beta = 0$ 
(the center of the prior in our post-reconstruction fits). 
In monopole, $\alpha_1$ works to change the position of un-weighted and $x$-weighted 
BAO features in opposite directions. 
The quadrupole changes sign when we switch $\alpha_1$ from $0.05$ to $-0.05$. 
The crest-trough feature at the BAO scale seen in the un-weighted and $x$-weighted quadrupoles 
move in opposite directions
with varying $\alpha_1$ analogous to the monopoles. }
\label{fig:b1effect}
\end{figure*} 

We may relate this parametrization to the $(\alpha, \epsilon)$ parametrization [or equivalently, $(\alpha_\perp,\alpha_\parallel)$] 
that have been used in recent BAO analyses \citep{Padmanabhan08Ani, AndersonDR11}.
In \cite{Padmanabhan08Ani}, the separation vectors between pairs of galaxies 
are parameterized by an isotropic dilation $\alpha(z)$ and an anisotropic warping $\epsilon(z)$ 
parameter. The deformation of the separation vector 
due to an incorrect distance redshift relation can be parameterized as 
\begin{align}
	r_\parallel &= \alpha(1+\epsilon)^2 r_\parallel^{\rm f}  \label{eq:para}\\ 
	r_\perp     &= \alpha(1+\epsilon)^{-1} r_\perp^{\rm f} \label{eq:perp} 
\end{align}
where the superscript ``f'' labels the fiducial values.
In the plane parallel limit, $r_\parallel = c\Delta z/H(z)$ and $r_\perp = \chi(z)\Delta \theta$. Here 
$\Delta z$ is the difference in redshifts of the two galaxies and $\Delta \theta$ is the angle
measured by the observer of the radial direction to each galaxy.
\begin{align}
\label{eq:pp_alpha}
	\alpha(z) &= \left[\frac{H_f(z)\chi^2(z)}{H(z)\chi_f^2(z)}\right]^{1/3}  \\
\label{eq:pp_epsilon}
	\epsilon(z) &= \left[\frac{H_f(z)\chi_f(z)}{H(z)\chi(z)}\right]^{1/3} - 1.
\end{align}
Together with Eq. \ref{eq:chi_ratio} and Eq. \ref{eq:h_ratio}, 
we can relate $\alpha(z)$ and $\epsilon(z)$ to $(\alpha_0,\alpha_1,\alpha_2)$. Working to linear order in
$\alpha_1$ and $\alpha_2$,
\begin{align}
\label{eq:alpha}
	\alpha(z) &=  \alpha_0\left[1+\frac{1}{3}\alpha_1+\frac{1}{3}(4\alpha_1+\alpha_2)x + \frac{5}{6}\alpha_2 x^2 \right] \\
\label{eq:eps}
	\epsilon(z) &= \frac{1}{3}\alpha_1 + \frac{1}{3}(\alpha_1+\alpha_2)x + \frac{1}{3}\alpha_2 x^2.
\end{align}

Fig. \ref{fig:b1effect} shows variations of the expected correlation function monopole
and quadrupole with $\alpha_1$ while holding $\alpha_0 = 1$ fixed. 
We have assumed a flat cosmology with 
$\Omega_m = 0.29$, $\Omega_b h^2 = 0.02247$, $h = 0.7$, $n_s = 0.97$, and $\sigma_8 = 0.8$
(the QPM cosmology described in Sec. \ref{sec:simulations}). 
In Fourier space, this model is given by the de-wiggled power spectrum as below in Eq. \ref{eq:dewiggled}. 
One can see from the monopole (left panel) that $\alpha_1$ causes shift of the un-weighted and $x$-weighed 
monopole BAO peaks in opposite directions. 
In contrast, since $\alpha_0$ causes isotropic shifts, it shifts the BAO peak 
in the un-weighted and $x$-weighted monopoles in the same direction. 
The quadrupoles (right panel) encode the anisotropic signal. 
Since Fig. \ref{fig:b1effect} assumes isotropic damping $\Sigma_\perp  = \Sigma_\parallel = 4.3 h^{-1}{\rm Mpc}$, 
the only anisotropic signal (quadrupole) comes from the 
mis-estimation of the distance-redshift relation characterized by $\alpha_1$. 
We see that the quadrupoles become inverted when we 
switch from $\alpha_1 = 0.05$ to $-0.05$. On top of the sign change, the BAO feature (the crest-trough at the acoustic scale) 
in the un-weighted and $x$-weighted quadrupoles shift in opposite directions analogous to the monopoles.

\subsection{Fitting the Correlation Function}

As in previous BAO analyses \citep{AndersonDR11}, we fit the galaxy correlation function
with a template. We describe this template below and discuss how it gets distorted due to a
mis-estimate of cosmology. 

In Fourier space, we use the following template for the 2D non-linear power spectrum \citep{Xu2013, AndersonDR11}
\begin{equation}
P_{t}\left(k,\mu\right)=\left(1+\beta\mu^{2}\right)^{2}F\left(k,\mu,\Sigma_{s}\right)P_{\text{dw}}\left(k,\mu\right).
\label{eq:pspec_template}
\end{equation}
The $\left(1+\beta\mu^{2}\right)^{2}$ term represents the Kaiser
effect \citep{Kaiser1987} with $\beta = f/b$ where $f\approx\Omega_{m}(z)^{0.55}$ \citep{Carroll92} is 
the growth rate of structure and $b$ is the large scale galaxy bias.
On small scales, the large random velocities in inner virialized clusters causes an elongation in the observed structure 
along the line-of-sight direction.
This is known as the Finger-of-God (FoG) effect and we model in Fourier space 
by the multiplicative factor $F\left(k,\mu,\Sigma_{s}\right)$
which takes the form
\begin{equation}
F\left(k,\mu,\Sigma_{s}\right)=\frac{1}{\left(1+k^{2}\mu^{2}\Sigma_{s}^{2}\right)^{2}}
\end{equation}
where $\Sigma_s$ is the streaming scale associated with the dispersion within clusters due to 
random peculiar velocities.

We model the degradation of the BAO due to non-linear structure growth by a Gaussian damping term. 
The damping is anisotropic due to redshift space distortions. 
The parallel and perpendicular streaming scales $\Sigma_{\parallel}$ and $\Sigma_{\perp}$ determine
 the amount of damping along and perpendicular to
the line-of-sight. The two streaming scales
are related by $\Sigma_\parallel = (1+f)\Sigma_\perp$ where $f$ is the 
growth rate of structure.
The de-wiggled power spectrum $P_{\text{dw}}\left(k,\mu\right)$ \citep{EisensteinSeoWhite}
is given by 
\begin{equation}
P_{\text{dw}}\left(k,\mu\right)=\left[P_{\text{lin}}(k)-P_{\text{nw}}(k)\right]\exp\left[-\frac{k_{\parallel}^{2}\Sigma_{\parallel}^{2}
+k_{\perp}^{2}\Sigma_{\perp}^{2}}{2}\right]+P_{\text{nw}}\left(k\right)
\label{eq:dewiggled}
\end{equation}
where $P_{\text{lin}}\left(k\right)$ is the linear power spectrum
from CAMB \citep{CAMB}. The no-wiggle spectrum $P_{\text{nw}}\left(k\right)$ is
the smoothed power spectrum \citep{Eisenstein&Hu} with the baryonic wiggles taken out. 

For our analyses, in pre-reconstruction fits, we fix
$\Sigma_s = 2 h^{-1}{\rm Mpc}$, $\Sigma_\perp = 6 h^{-1}{\rm Mpc}$ and $\Sigma_\parallel = 9.6 h^{-1}{\rm Mpc}$. 
For post-reconstruction, we use
$\Sigma_s = 0 h^{-1}{\rm Mpc}$, $\Sigma_\perp = \Sigma_\parallel = 4.3 h^{-1}{\rm Mpc}$.
These prescribed parameters are motivated by fitting to the average mock correlation function of the mocks
we use. Before reconstruction, the difference in the streaming parameters $\Sigma_\perp$ and $\Sigma_\parallel$ 
come from the Kaiser effect. Reconstruction is expected to remove the Kaiser squashing, and hence our choice 
of $\Sigma_\parallel = \Sigma_\perp$ after reconstruction.
In the fits to the average correlation function, the streaming parameter $\Sigma_s$ is not well-constrained. 
However, we have checked that fitting the BAO feature in individual mocks is insensitive 
to the choice of these streaming parameters around our prescribed values. 

The multipole moments of the template power spectrum can be computed
as 
\begin{equation}
P_{\ell,t}(k)=\frac{2\ell+1}{2}\int_{-1}^{1}P_{t}\left(k,\mu\right)L_{\ell}(\mu)d\mu
\end{equation}
where $L_\ell$ is the Legendre polynomial of order $\ell$. 
To calculate the correlation functions, we Fourier transform the power spectrum as
\begin{equation}
\xi_{\ell,t}(r)=i^{\ell}\int\frac{k^{3}d\log k}{2\pi^{2}}P_{\ell,t}(k)j_{\ell}(kr).
\end{equation}

Now we review how a misestimate of the cosmology distorts the correlation function. 
A perturbative expression is given by Eq. 26 and 27 in \cite{Xu2013}. 
However, we use a different  
approach here. With Eq. \ref{eq:para} and \ref{eq:perp}, we can express
the true galaxy separation and the cosine of the angle between the separation vector and line-of-sight 
in terms of the fiducial values by using $\alpha$ and $\epsilon$. Given 
\begin{align}
	r      &= \sqrt{r_\parallel^2 + r_\perp^2}, \\
	\mu &= \cos\left[\arctan\left(\frac{r_\perp}{r_\parallel}\right)\right],
\end{align}
we get,
\begin{align}
	r      &= \alpha r^{\rm f}\sqrt{(1+\epsilon)^4 (\mu^{\rm f})^2 + (1+\epsilon)^{-2}[1-(\mu^{\rm f})^2]} \\
	\mu &= \cos[\arctan[(1+\epsilon)^{-3}\tan(\arccos \mu^{\rm f})]].
\end{align}
These are the ``true'' separation and line-of-sight angle that go into the true correlation function, which  
can be decomposed into multipole moments using the Legendre Polynomials 
\begin{equation}
	\xi_t(r,\mu) = \sum_\ell \xi_{\ell,t}(r) L_\ell(\mu)
\end{equation} 
where we ignore the contributions from $\ell=10$ or higher. 
We find the expansion to be quickly converging and 
the amplitudes of higher order multipoles are significantly reduced. 
A non-linear model for $\xi_{\ell,t}(r)$ is given in the next subsection.

Substituting $r$ and $\mu$ with the expressions above, we reach 
the model correlation function $\xi(r^{\rm f}, \mu^{\rm f}, \alpha, \epsilon)$. 
This model correlation function includes the ``isotropic dilation" and ``anisotropic warping" 
due to incorrectly assuming a fiducial cosmology. 
We then re-project onto Legendre polynomials
\begin{equation}
	\xi_{\ell,m}(r, z) = \frac{2\ell +1}{2}\int_{-1}^1 \xi(r^{\rm f},\mu^{\rm f},\alpha,\epsilon)L_\ell(\mu^{\rm f})d\mu^{\rm f}.
\label{eq:alpha_epsilon_xi}
\end{equation}
This is our template for matter correlation function within a redshift slice.

\subsection{Redshift Weights}

We define weights to compress the information in the redshift direction
onto a small number of ``weighted correlation functions". 
The weights are designed to optimally extract the constraints on $\alpha_0$, $\alpha_1$, $\alpha_2$. 
We refer the reader to \cite{Zhu15} for the derivation of the weights which are modeled on \cite{TTH}. 
The weights constructed for the distance-redshift parametrization in Sec. \ref{sec:distance_redshift}
are given by a multiplicative quantity $w_{\ell,i}d{\cal W}$. Here, $d{\cal W}(z)$ is given by 
\begin{equation}
\label{eq:dW}
	d{\cal W}(z) = \left(\frac{\bar{n}}{\bar{n}P+1}\right)^2 dV(z)
\end{equation}
where the volume of the slice is given by
\begin{equation}
	dV(z) = \frac{\chi_f^2(z)}{H_f(z)} dz d\Omega .
\end{equation}
$d{\cal W}(z)$ is the inverse of the variance
of the correlation function bin at redshift $z$. We assume that different redshift bins 
are independent, so that 
the covariance matrix across redshifts is diagonal. In the equation above, $P$ is the power 
at the BAO peak scale, and is specified to be $10^4 h^{-3}{\rm Mpc}^3$.

The additional weights $w_{\ell,i}$ are given by
\begin{align}
%%%%%
w_{0,\alpha_0} &=& 1 & &
w_{2,\alpha_0} &=& 0 \\
%%%%%
w_{0,\alpha_1} &=& \frac{1}{3}(1+4x) & &
w_{2,\alpha_1} &=& \frac{1}{3}(1+x) \\
%%%%%
w_{0,\alpha_2} &=& \frac{1}{6}(2x+ 5x^2) & &
w_{2,\alpha_2} &=& \frac{1}{3}(x+x^2)
\end{align}
The first indices $\ell = 0, 2$ indicate the weights are for fitting the monopole or quadrupole 
moments of the correlation function. The second indices $\alpha_i$ indicate the parameter
one is focusing on.
%Notice that the weights are linear combinations of 1, $x$, and $x^2$. 
%In practice, we can fit the correlation functions
%weighted by $w_{\ell,i}$ or by weighted by $1$, $x$, and $x^2$. 
%The analysis in this paper uses  the latter for convenience.  

\section{Simulations}
\label{sec:simulations}

We test our algorithm
on mock galaxy catalogs created by using the ``quick particle mesh" (QPM)
method \citep{WhiteQPM}. These catalogs are constructed
to simulate the clustering and noise level of the SDSS DR12 combined samples. 
For details of BOSS survey design, we refer the reader to \cite{EisensteinDesign} and 
\cite{Dawson13}.
The mock catalogs are based on 1000 low force- and mass-resolution
particle-mesh N-body simulations. Each uses $1280^3$ particles in a box of side length $2560 h^{-1}{\rm Mpc}$. 
The simulations assume a flat $\Lambda$CDM cosmology, with cosmological parameters as
 : $\Omega_m = 0.29$, $\Omega_b h^2 = 0.02247$, $h = 0.7$, $n_s = 0.97$, and $\sigma_8 = 0.8$. 
These mocks are constructed from 1000 QPM realizations, each of which starts at $z = 25$ using 
second order Lagrangian perturbation theory.
The catalogs span the redshift range of $z=0.2$ to 0.7 and cover both the northern and southern Galactic cap of
the BOSS footprint. The mocks are populated using a bias model inferred from small scale measurements, 
and have a redshift dependent galaxy bias reflecting changes in the galaxy population over the BOSS redshift range.
The mocks include the effects of the angular veto mask of the BOSS galaxies, as well as an approximation to fiber collisions. The redshift selection
function $n(z)$ was matched to the angular density of the DR12 sample, to make it independent of cosmology.

%Veto mask has been applied to all mocks. FKP weights are added to both mocks and randoms. 
%Fiber collisions are approximated in these mocks. The selection function 
%$n(z)$ matches the angular density of the DR12 sample rather than space density, 
%to make this match independent of cosmology.

%
%\subsection{Fiducial Cosmologies}
%We use two fiducial cosmologies in our analyses. We will conduct the analysis with
%both and argue the measured distance-redshift relation measured from the fitting results remain the same
%in both cases. 
%
%The first is the same cosmology the simulation adopts. 
%Under this fiducial cosmology, one is expected to measure $\langle \alpha_0 \rangle = 1$ and $\langle \alpha_1 \rangle  = 0$.
%
%The second cosmology assumes 
%$\Omega_M = 0.25$, and $\Omega_M h^2 = 0.1421$ is kept the same as QPM cosmology. 
%This determines the Hubble constant $h = 0.754$. $\Omega_b h^2 = 0.02247$ is also 
%kept to be the same to ensure
%the sound horizon stays the same in comoving units.
%

\section{Analysis}
\label{sec:analysis}

\subsection{Computing the weighted correlation functions}

We analyze the simulations similar to previous BOSS analyses \citep{AndersonDR11}. We refer the reader to those papers for more
detailed descriptions, restricting our discussion to the new aspects. 
The first of these is that we treat the entire BOSS redshift range
as a unified sample (from $z=0.2$ to $0.7$) and do not split into smaller redshift bins. 
Since the efficacy of the BAO reconstruction
procedure has now been well established and our redshift weights are agnostic to reconstruction, 
our default results will all be post-reconstruction.
Our implementation of reconstruction is identical to what has been used for the SDSS and BOSS analyses \citep{AndersonDR11}.

In order to compute the weighted correlation functions, 
we use a modified version of the Landy-Szalay estimator \citep{LS}.  
As is traditional, we weight every galaxy/random by the FKP weight
\begin{equation}
	w_{\rm FKP} = \frac{1}{1+\bar{n}(z) P(k_0)}
\end{equation}
where $\bar{n}(z)$ is the number density at $z$, the redshift of the object. 
$P(k_0) = 10^4 h^{-3}{\rm Mpc}^3$ is the approximate power at the BAO scale. 

We also weight each {\bf pair} of galaxies/randoms by $w = 1$, $x$, $x^2$ to construct the weighted correlation
functions $\xi_1$, $\xi_x$ and $\xi_{x^2}$. 
Since the redshift separation between a pair that contributes to the correlation function is small, 
we simply use the mean redshift of each pair to compute $x$. 
The weighted 2D correlation functions are then given by
\begin{equation}
\label{eq:weighted_LS}
	\xi_w^{\rm data}(r,\mu)=\frac{\widetilde{DD}(r,\mu) - 2 \widetilde{DR}(r,\mu) + \widetilde{RR}(r,\mu)}{RR(r,\mu)}
\end{equation}
where $\widetilde{DD}$, $\widetilde{DR}$ and $\widetilde{RR}$ include the additional pair weight, 
whereas $RR$ in the denominator does not. After reconstruction, this gets modified to
\begin{equation}
\label{eq:weighted_LS_reconstructed}
	\xi_w^{\rm data}(r,\mu)=\frac{\widetilde{DD}(r,\mu) - 2 \widetilde{DS}(r,\mu) + \widetilde{SS}(r,\mu)}{RR(r,\mu)}
\end{equation}
where $S$ represents the shifted random particles. 

In computing the pair sums, we bin the weighted pair sums in both $r$ and $\mu$. 
The $r$ bins used here are from 0 to 200 with 4 $h^{-1}{\rm Mpc}$
bins. The $\mu$ bins are from 0 to 1 with 0.01 in width. 
From the 2D correlation function, one can compute the monopole and quadrupole moments as
\begin{equation}
	\xi_{\ell,w}^{\rm data}(r) = \frac{2\ell + 1}{2} \int_{-1}^1 \xi_w^{\rm data}(r,\mu) L_\ell(\mu) d\mu
\label{eq:xi_data}
\end{equation}
where $L_\ell$ is the Legendre polynomial of order $\ell$. 

%In practice, the correlation functions we measure are binned. 
%The correlation functions are rebinned to coarser (i.e. 4Mpc/h) resolution from the original 1Mpc/h resolution.
We bin our estimators accordingly to the 4 $h^{-1}{\rm Mpc}$ resolution.
\begin{equation}
\xi_\ell(r_{\rm cen}) = \frac{3}{r_2^3 - r_1^3} \int_{r_1}^{r_2} r^2 \xi_\ell(r)dr
\end{equation}
gives the binned correlation function. The bin is centered at $r_{\rm cen}$, with a lower bound $r_1$ and an upper bound $r_2$.

\subsection{Weighted Correlation Function Estimators}

We construct models of the monopoles and quadrupoles of the unweighted and weighted 
correlation functions. 
Since the additional weights $w_{\ell,i}$ all take the simple form of linear combinations of 1, $x$, and $x^2$, 
it is convenient to calculate
correlation functions weighted by them instead of the original weights.
Using these weights, the weighted correlation function estimators
can be constructed as weighted integrals,
\begin{align}
	\xi_{\ell,1}(r) & =  \frac{1}{N}\int  d\mathcal{W}(z) b^2(z) \xi_{\ell,m}(r, z)\\
	\xi_{\ell,x}(r) & =  \frac{1}{N}\int d\mathcal{W}(z)x(z) b^2(z)  \xi_{\ell,m}(r, z)\\
	\xi_{\ell,x^{2}}(r) & =  \frac{1}{N}\int d\mathcal{W}(z)x^{2}(z) b^2(z) \xi_{\ell,m}(r, z)
\end{align}
where $N=\int d\mathcal{W}$ is a convenient choice of normalization and $b(z)$ is the galaxy bias.
We assume that the bias is inferred from small-scale clustering measurements. 
We demonstrate that our results are robust to small changes in input form of $b(z)$.
The above integrals are understood 
to be over the redshift range of the survey. 

It is more efficient to compute the weighted integrals as summations 
across redshifts. 
To do this, we bin the redshift range of the combined sample [0.2, 0.7] 
into 50 thinner slices of width $\Delta z = 0.01$. 
We use the central redshift of each slice to label these slices. 
In each redshift bin, with the given parameters $\alpha_0$, $\alpha_1$, and $\alpha_2$, 
one calculates $\chi(z)/\chi_f(z)$ and $H_f(z)/H(z)$ 
according to 
Eq. \ref{eq:chi_ratio} and Eq. \ref{eq:h_ratio}. 
Using the obtained $\chi$ and $H$ ratios in Eq. \ref{eq:pp_alpha} and Eq. \ref{eq:pp_epsilon}, one calculates the 
``isotropic dilation" parameter $\alpha(z)$ and ``anisotropic warping" parameter $\epsilon(z)$ at different redshifts.
Alternatively, one can directly use Eq. \ref{eq:alpha} and Eq. \ref{eq:eps} to get $\alpha(z)$ and $\epsilon(z)$. 
This feature is distinct from traditional analyses in which $\alpha$ and $\epsilon$ are only measured at the 
``effective" redshift of the sample. 

For efficient calculation of the redshift dependent $\xi_{\ell,m}(r, z)$,
we pre-compute and tabulate correlation function monopoles and quadrupoles by
fixing $\alpha = 1$ while $\epsilon$ ranges from 
-0.2 to 0.2 with intervals of 0.001.  
We first calculate the correlation function by interpolating in the $\epsilon$ direction. 
This gives us the correlation function corresponding to $\alpha = 1$ and $\epsilon = \epsilon(z)$. 
We then interpolate the obtained correlation function at separation scale $\alpha(z) r$.

Within each slice, we also calculate the inverse variance factor $\Delta {\cal W}(z)$ and the additional weights $x(z)$ 
as they will be used to 
weight the correlation functions in different slices. In calculating $\Delta {\cal W}(z)$ using Eq. \ref{eq:dW}, 
the volume of each slice is calculated as 
\begin{equation}
	\Delta V(z) \propto \frac{\chi_f^2(z)}{H_f(z)} \Delta z.
\end{equation}

Once all these ingredients are in hand, we weight the correlation function monopoles and quadrupoles
 by $\Delta {\cal W}(z) w_z$ where $w_z = 1,x,x^2$ and sum across redshifts.
 We thus achieve the ``un-weighted", ``$x$-weighted", and ``$x^2$-weighted" monopole and quadrupole estimators. 

%
%\subsection{Reconstruction}
%
%Non-linear structure formation smears the acoustic peak and degrades the sensitivity of the BAO measurements. 
%To partially reverse the undesirable nonlinear effects, 
%we apply the density field reconstruction technique suggested in \citet{EisensteinRecon07}. 
%The technique is based on mapping the galaxy displacement field from the observed galaxy density field by using the 
%Lagrangian perturbation theory. The galaxies are then 
%moved back to where they would have been in linear theory. 
%before we calculate the pair counts. 
%Reconstruction sharpens the acoustic peak and improves the precision of BAO scale measurements. It also
%removes the shift of the BAO peak due to non-linear effects. 
%
%We apply reconstruction to our mock galaxy catalogs. We use the galaxy density field to estimate the underlying
%matter density field and solve for the displacement field. More details of the algorithm can be found in 
%\citet{Padmanabhan2012}. 
%

\subsection{Fitting the Acoustic Feature}

The fitting aims to minimize the $\chi^2$ goodness-of-fit indicator given by 
\begin{equation}
	\chi^2 = (\vec{m} - \vec{d})^T C^{-1} (\vec{m} - \vec{d}).
\label{eq:chi2}
\end{equation}
We describe the data vector $\vec{d}$, the model vector $\vec{m}$,
and the covariance matrix $C$ below.

We perform two sets of fits on the mock correlation functions with the model outlined in
the previous sections.
In the first set of fits, we fit the ``unweighted'' correlation functions. 
We will call this set of fits ``unweighted fits" or ``1 fits".
In the second set, we simultaneously fit the unweighted and the $x$-weighted correlation functions.
We will call this set ``weighted fits" or ``1+x fits". 

We adopt $48 h^{-1}{\rm Mpc} < r < 152 h^{-1}{\rm Mpc}$ as our fiducial fitting range with $4 h^{-1}{\rm Mpc}$ bins. 
We use the bin center to label each bin. The monopole and quadrupole data vector $\vec{d}_{\ell,w}$
corresponds to 26 points each, with $50 h^{-1}{\rm Mpc}$ being the first bin and $150 h^{-1}{\rm Mpc}$ the last one.

For ``unweighted" fits, we simultaneously fit the unweighted monopole and quadrupole correlation function 
$\vec{d}_{0,1}$ and $\vec{d}_{2,1}$.
The data vector and model vector take the form
\begin{equation}
\vec{d}=\left(\begin{array}{c}
\vec{d}_{0,1}\\
\vec{d}_{2,1}
\end{array}\right)\quad\vec{m}=\left(\begin{array}{c}
\vec{m}_{0,1}\\
\vec{m}_{2,1}
\end{array}\right).
\end{equation}
The monopole/quadrupole are denoted by $\ell=0,2$ respectively, while $w=1,x$ indicate the $z$-weight.
%We denote the model vector of the same size as $\vec{m}_{\ell,w}$.

For the ``weighted" fits (``1+x"), we simultaneously fit the unweighted and $x$-weighted monopoles and quadrupoles. 
The data vector and model vector take the form
\begin{equation}
	\vec{d}=\left(\begin{array}{c}
	\vec{d}_{0,1}\\
	\vec{d}_{2,1}\\
	\vec{d}_{0,x}\\
	\vec{d}_{2,x}
	\end{array}\right)
	\quad
	\vec{m}=\left(\begin{array}{c}
	\vec{m}_{0,1}\\
	\vec{m}_{2,1}\\
	\vec{m}_{0,x}\\
	\vec{m}_{2.x}
\end{array}\right)
\end{equation}
The data vectors $\vec{d}_{\ell,w}$ are given by $\xi_{\ell,w}^{\rm data}(r)$ in Eq. \ref{eq:xi_data}. 
The model vectors $\vec{m}_{\ell,w}$ 
will be explained in detail in the next subsection. 
Once again, in the combined column vector $\vec{d}$ and $\vec{m}$, each vector $\vec{d}_{\ell,w}$ and $\vec{m}_{\ell.w}$
corresponds to 26 points. 

\subsubsection{The Fiducial Fitting Model}

We fit our correlation functions to
\begin{equation}
	\xi^{\rm fit}_{\ell,w}(r) = B_w^2 \xi_{\ell,w}(r) + A_{\ell,w}(r)
\end{equation}
where $\xi_{\ell,w}(r)$ is the weighted correlation function while
$A(r)$ absorbs un-modeled broadband
features including redshift-space distortions and scale-dependent bias
following \cite{AndersonDR11}. We assume
\begin{equation}
A_{\ell,w}\left(r\right)=\frac{a_{\ell,w,1}}{r^{2}}+\frac{a_{\ell,w,2}}{r}+a_{\ell,w,3}.
\end{equation}
We allow a multiplicative factor $B_w^2\sim1$ to vary in order to adjust the amplitudes of the correlation functions. 
Note that $B_w^2$ determines the amplitudes of the monopole and quadrupole together while $\beta$ sets the relative 
amplitude between the two. 

\subsubsection{Covariance Matrices}
\label{sec:covariance}
The most direct way to calculate the covariance matrix is from the mock catalogs.
The $(i,j)$ element of the covariance matrix is calculated as 
\begin{equation}
C_{ij}=\frac{1}{N_s-1}\sum_{n=1}^{N_s}\left[d_{n}\left(r_{i}\right)-\bar{d}\left(r_{i}\right)\right]\left[d_{n}\left(r_{j}\right)-\bar{d}\left(r_{j}\right)\right]
\end{equation}
where $N_s$ is the total number of mocks, $d_n(r)$ is the correlation function calculated from the $n$th mock
and $\bar{d}(r)$ is the average of the mock correlation functions. 

When estimating the inverse covariance matrix, $\Psi$, from mocks, we account for the bias from the asymmetry of the 
Wishart distribution by multiplying the inverse covariance matrix by a prefactor $(1-D)$, namely, 
$\Psi = (1-D)C^{-1}$ \citep{Hartlap07, PercivalCovar} where
\begin{equation}
	D = \frac{N_b + 1}{N_s - 1}.
\end{equation}
Here $N_b$ is the size of the data vector. 

This correction is also important in calculating the expected $\chi^2$ value.  
If one is fitting a sample by using the covariance matrix calculated from the same sample,
the expected $\chi^2$ is equal to the degree-of-freedom multiplied by the prefactor $(1-D)$.
We refer the reader to the appendix for a derivation of this relation.

\subsubsection{Summary of Parameters}

In the unweighted fits, the non-linear parameters we fit for are 
$B_1^2, \beta, \alpha_0$, and $\alpha_1$, in addition to the $2 \times 3 = 6$ 
linear nuisance parameters in $A_{\ell,w}(r)$, a total of 10 parameters. Note that 
$\ell = 0,2$ and $w = 1$, yielding a data vector with 52 elements,
and a fit with 42 degrees of freedom. 
We calculate the expected $\chi^2$ for individual mocks by including the prefactor 
described in Sec. \ref{sec:covariance}, and get the expected $\chi^2$ to be $40$. 

Similarly, in the weighted fits, the non-linear parameters we fit for are 
$B_1^2, B_x^2, \beta, \alpha_0$, and $\alpha_1$, in addition to the $4\times 3 = 12$
linear nuisance parameters in $A_{\ell,w}(r)$ where $w = 1$ or $x$. This gives a total of
17 parameters of interest. Therefore, ${\rm dof} = 4\times 26 - 17 = 87$ in the weighted fit. 
This yields the expected $\chi^2$ to be $78$. 

We obtain the set of best-fit model parameters by minimizing $\chi^2$ as in Eq. \ref{eq:chi2}. 
The non-linear parameters are handled through a simplex algorithm \citep{nelder1965simplex} while
the linear nuisance parameters are obtained using a least-squares method nested within the simplex. 
For each set of non-linear parameters, the least-squares algorithm gives the corresponding best-fit linear 
parameters. The simplex algorithm then searches the non-linear parameter space until the best-fit parameters 
that minimize $\chi^2$ are achieved . 

Some mocks possess a weak BAO feature. The low signal-to-noise causes the nuisance polynomial to 
become the dominant contribution to the model correlation function. 
To avoid these undesirable cases, 
we adopt a Gaussian prior for $\beta$ centered around 0.35 with standard deviation 0.15. 
%we also adopt a Gaussian prior on $\log(B_1^2)$ and $\log(B_x^2)$ at 0 with standard deviation 0.4. 
After reconstruction, we put a Gaussian prior of the same width centered around 0 as 
reconstruction partially removes the Kaiser effect.

In the default fits, we allow $\alpha_0$ and $\alpha_1$ to float while fixing $\alpha_2 = 0$.
We discuss extending our fits to include $\alpha_2$ in Sec. \ref{sec:including_a2} below.

\section{Results}
\label{sec:results}

\subsection{Fiducial Results}

We present the results of the fits to the QPM mock correlation functions using
both the ``unweighted" and the ``weighted" fits. The fits assume the QPM 
cosmology as the fiducial cosmology and assume a pivot redshift $z_0 = 0.57$. 
We will then compare the results from ``unweighted'' and ``weighted" fits 
and comment on the effectiveness of redshift weighting in measuring
the distance-redshift relation and the Hubble parameter
to a higher accuracy. 

\begin{figure*}
\centering \includegraphics[width=\textwidth]{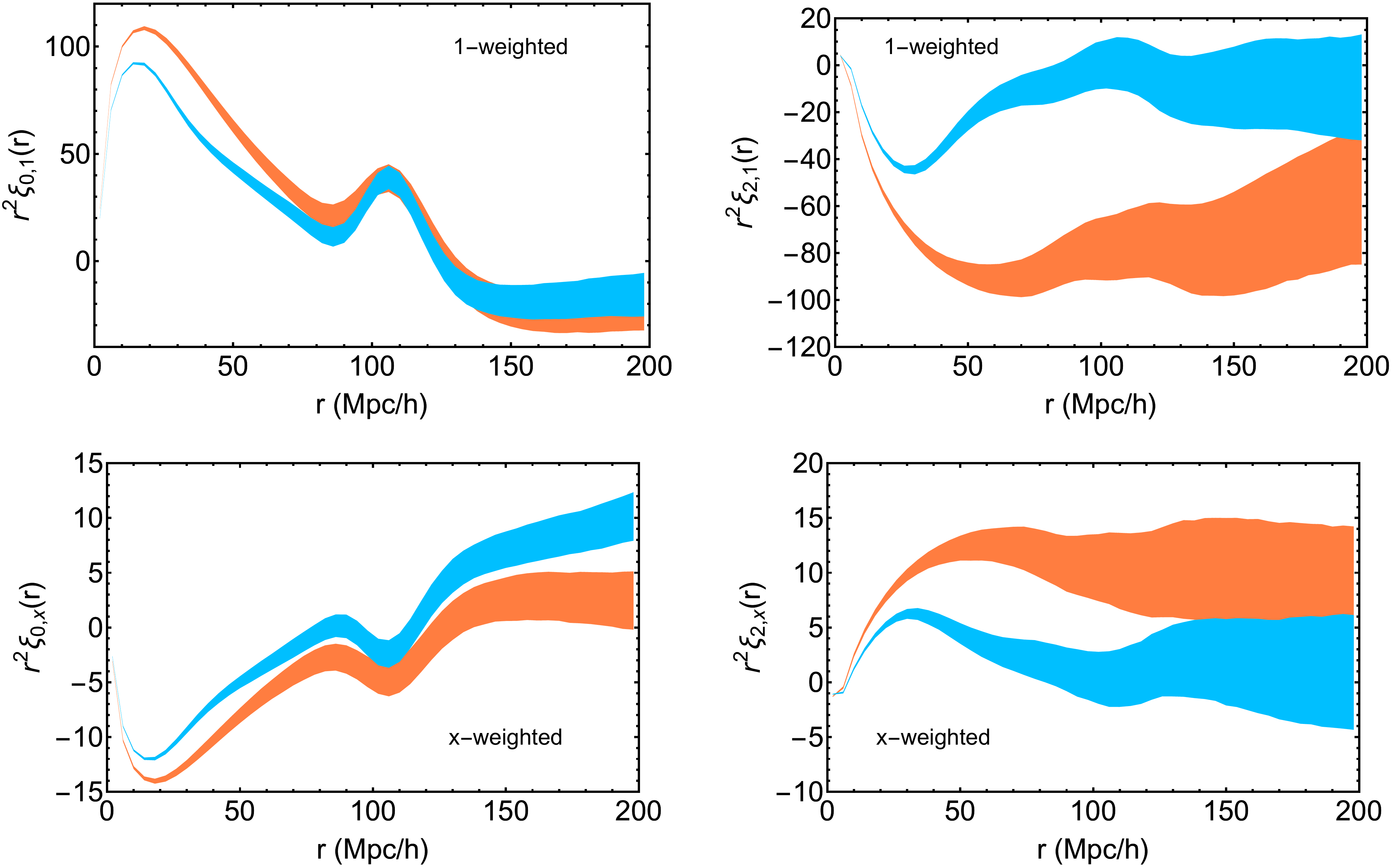}
\caption{The average monopoles (left) and quadrupoles (right) from the 1000 QPM mocks. 
The error bands plotted are that of an individual mock, which are $\sqrt{1000}$ bigger than for the that of the 
average correlation function. The orange bands are pre-reconstruction mocks, 
while the blue bands correspond to post-reconstruction. 
The top panels show the ``unweighted" monopoles and quadrupoles while the bottom show the ``$x$-weighted" 
ones. 
One can see that the ``$x$-weighted" monopoles and quadrupoles are inverted as compared to the ``unweighted" ones, due to an
overall negative weight. 
The acoustic feature is clearly visible in the ``$x$-weighted" monopoles. 
The reconstructed monopole moments show a sharpened acoustic peak, suggesting reconstruction partially 
removes the degradation of BAO due to non-linear evolution. 
The quadrupole amplitudes are significantly reduced after reconstruction. At large scales,
quadrupole moments are close to 0, indicating the efficiency of reconstruction at 
removing the Kaiser effect.
 }
\label{fig:average}
\end{figure*}

We plot the average monopole and quadrupole of 1000 mocks before and after reconstruction in Fig. \ref{fig:average}.
The bands contain the error for individual mocks. 
One can see that the ``$x$-weighted" monopoles and quadrupoles are inverted as compared to the ``unweighted" ones.
The inversion comes from an overall negative weight. 
Albeit inverted, the acoustic feature is clearly visible in the ``$x$-weighted" monopoles. 
A comparison of the monopoles before and after reconstruction shows that the acoustic peak in the monopole is 
more pronounced after reconstruction, suggesting reconstruction is effective in partially undoing the damping of the BAO feature
due to nonlinear evolution. 
Motivated by a fit to the average correlation function, we have chosen $\Sigma_\perp = 6 h^{-1}{\rm Mpc}$ and $\Sigma_\parallel = 9.6 h^{-1}{\rm Mpc}$ before reconstruction
and $\Sigma_\perp = \Sigma_\parallel = 4.3 h^{-1}{\rm Mpc}$ for post-reconstruction fits.
In addition, one can see the quadrupole amplitude is substantially smaller and close to zero after reconstruction on large scales. 
This confirms that reconstruction partially removes the effects of redshift space distortion. 

\begin{figure*}
\centering 
\includegraphics[width=\textwidth]{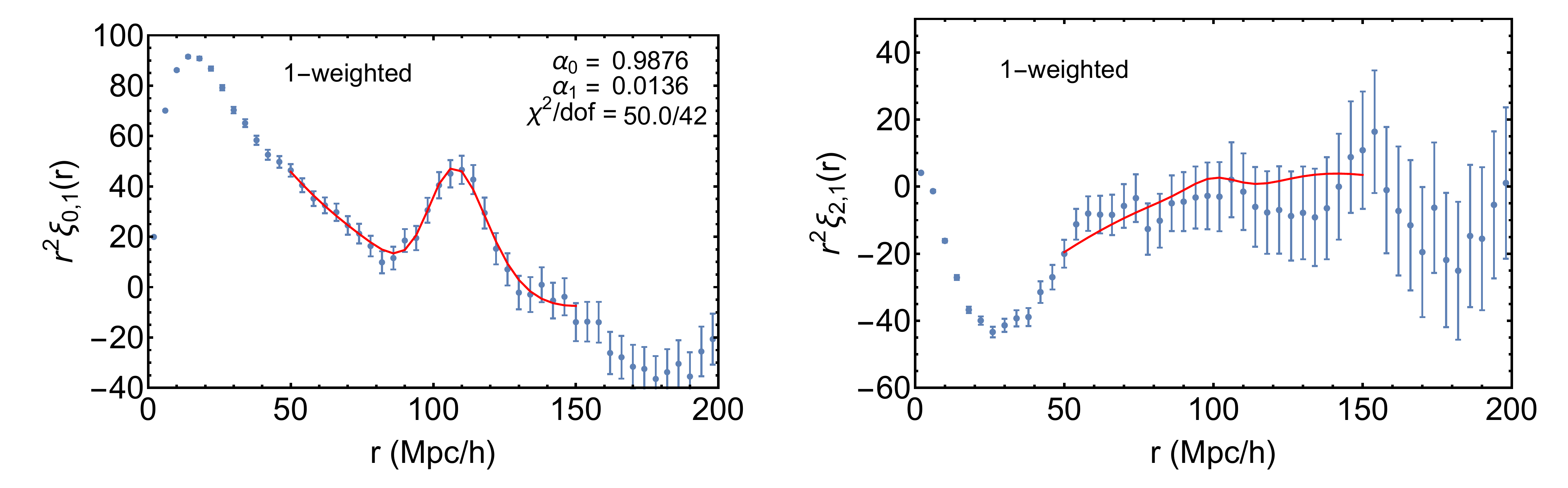}
\caption{The best-fit to a sample unweighted monopole (left) and quadrupole (right) from a reconstructed mock 
correlation function. 
The [blue] bars are monopoles and quadruples from a sample mock with $1\sigma$ error bars. 
The solid [red] lines are the best-fit models.  
The best fit parameters and the corresponding $\chi^2$ values are listed on the figure.}
\label{fig:1_sample}
\end{figure*}

\begin{figure*}
\includegraphics[width=\textwidth]{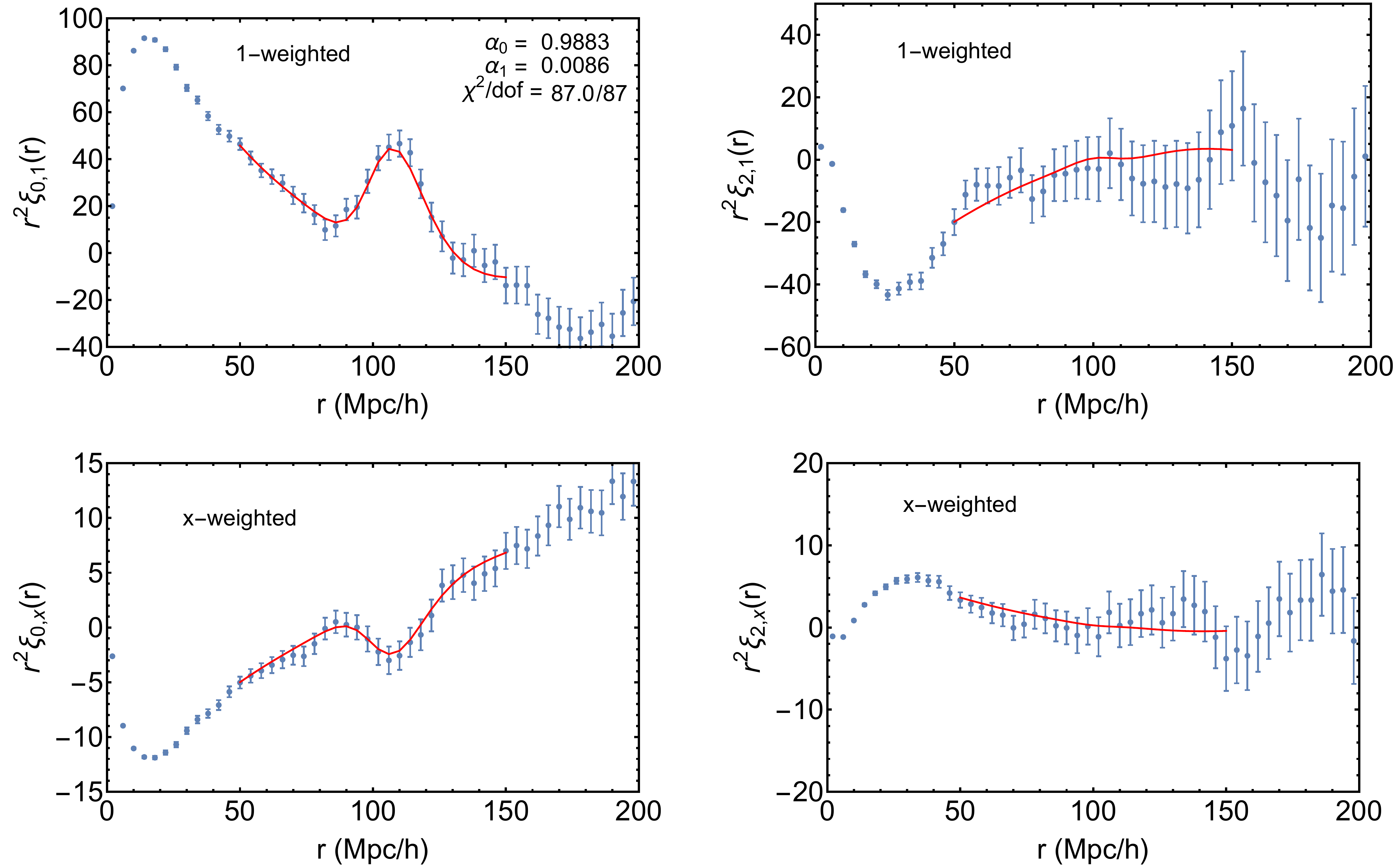}
\caption{ Sample ``weighted" fit to the unweighted and $x$-weighted monopoles (left) and quadrupoles (right) of the same mock 
as in Fig. \ref{fig:1_sample}. 
The top two panels are the unweighted monopoles and quadrupoles while the bottom two are weighted by $x$. 
The best-fit models (solid [red] lines) are over-plotted.  }
\label{fig:1+x_sample}
\end{figure*}

We measure $\alpha_0$ and $\alpha_1$ for each mock using the fitting procedure and model outlined in \S \ref{sec:analysis}. 
Since our fiducial cosmology is the same as simulation cosmology, 
we expect $\langle \alpha_0\rangle = 1$ and $\langle \alpha_1 \rangle = 0$ if our estimators are unbiased.
Fig.~\ref{fig:1_sample} shows fit to an example ``unweighted" post-reconstruction 
monopole and quadrupole, 
while Fig.~\ref{fig:1+x_sample} shows the ``weighted" fit to the same mock where we simultaneously 
fit the ``unweighted" and ``$x$-weighted" monopoles and quadrupoles. 

%
%In the pre-reconstruction case, the ``unweighted" fits yield $\langle \alpha_0 \rangle = 1.0022 \pm 0.0118$ and 
%$\langle \alpha_1 \rangle = 0.0036 \pm 0.0478$. 
%After applying the $z$-weights, we measure $\langle \alpha_0 \rangle = 1.0023 \pm 0.0115$. 
%We see that the error on $\alpha_0$ is comparable to the value from the ``unweighted" fits.
%The $z$-weights reduce the error bar on $\alpha_1$ and yield $\langle \alpha_1 \rangle = 0.0020 \pm 0.0332$. 
%We see a 30\% decrease in the error of $\alpha_1$ by using the optimal $z$-weights. 
%Note that these measurements are all consistent with $\alpha_0$ and $\alpha_1$ to be centered around 1 and 0
%within uncertainties. It indicates that redshift weighting gives non-biased measures of $\alpha_0$ and $\alpha_1$. 
%

\begin{table*}
\centering
\begin{tabular}{ l | c | c | c  }
    \hline
     Model &  $\alpha_0$ & $\alpha_1$ & $\langle \chi^2 \rangle/{\rm dof}$\\ \hline
     \hline
      & Before Reconstruction & & \\ \hline
      \hline
      Fiducial, weighted  & $1.0023 \pm 0.0115$ & $0.0020 \pm 0.0332$  & $78.66/87$ \\ \hline
      Fiducial, unweighted & $1.0022 \pm 0.0118$ & $0.0036 \pm 0.0478$ & $40.53/42$ \\ \hline
    \hline
    & After Reconstruction & & \\ \hline
    \hline
    $\textbf{Fiducial, weighted}$  &    $\bf{1.0005 \pm 0.0079}$ & $\bf{0.0025 \pm 0.0205}$  & $\bf{80.00/87}$\\ \hline
    Fiducial, unweighted  & $1.0002 \pm 0.0084$ & $0.0050 \pm 0.0276$ & $41.87/42$ \\ \hline
     Fit w/  $(\Sigma_\perp,\Sigma_\parallel) \to (3,3) h^{-1}{\rm Mpc}$ & $1.0002\pm 0.0079$ & $0.0030\pm 0.0207$ & $80.61/87$ \\ \hline
     Fit w/ $\Sigma_s = 2 h^{-1}{\rm Mpc}$ & $1.0004\pm 0.0079$ & $0.0036\pm 0.0205$ & $80.30/87$ \\ \hline
     Fit w/ constant $b(z) = 1.7$ & $1.0002\pm 0.0079$  & $0.0030\pm 0.0209$ & $80.00/87$\\ \hline
       Fit w/ $70 < r < 150 h^{-1}{\rm Mpc}$ & $1.0004\pm 0.0079$ & $0.0021\pm 0.0208$ & $61.22/67$ \\ \hline
       $z_{\rm pivot} = 0.4$ & $1.0000\pm 0.0102$ & $0.0015\pm 0.0135$ & $ 80.01/87 $ \\ \hline
       Fit w/ $\Omega_m = 0.25$ cosmology, assuming $\alpha_2 = 0$ & $1.0603\pm 0.0084$ & $-0.0145 \pm 0.0203$ &  $80.30/87$ \\  \hline
       Fit w/ $\Omega_m = 0.25$ cosmology with floating $\alpha_2$ \\ (expect $\alpha_0 = 1.0599$, $\alpha_1 = -0.0161$, and $\alpha_2 = 0.0018)$ & $1.0605\pm 0.0083$ & $-0.0155 \pm 0.0205$ &  $119.97/131$ \\ 

    \hline
\end{tabular}
\caption{Mean and standard deviations of best-fit $\alpha_0$ and $\alpha_1$ from  
``unweighted fits" and ``weighted fits" with various models. The model is given in column 1. 
The mean and standard deviation of the best-fit parameters from the mocks are given in column 2 and 3. 
The mean $\chi^2/{\rm dof}$ is given in column 4. 
For a relation between the expected average $\chi^2$ and {\rm dof}, see the appendix.
%Note that $\alpha_0 = 1$ and $\alpha_1 = 0$ in the QPM cosmology
%correspond to $\alpha_0 = 1.0599$ and $\alpha_1 = -0.0161$ in the $\Omega_m = 0.25$ cosmology.
}
\label{tab:bestfit_tab}
\end{table*}

A summary of our fitting results is in Table~\ref{tab:bestfit_tab}. 
The results are all consistent with expected values within uncertainties, 
suggesting our weighted correlation functions are an unbiased estimator. 
Furthermore, applying the $z$-weights significantly reduce the $\alpha_0$ and $\alpha_1$ errors. 

%
%After reconstruction, the error on both $\alpha_0$ and $\alpha_1$ is reduced. This is expected as reconstruction
%makes the acoustic peak less smeared, allowing a better measurement of the peak position. 
%For $\alpha_0$, the ``unweighted" fits yield $\langle \alpha_0 \rangle = 1.0002 \pm 0.0084$ and 
%we measure $\langle \alpha_0 \rangle = 1.0005 \pm 0.0079$. A comparison with the pre-reconstruction case suggests that
%the mean best-fit value of $\alpha_0$ moves closer to 1, suggesting that reconstruction removes some effects of 
%non-linear structure growth. 
%
%For $\alpha_1$, the ``unweighted" fits yield $\langle \alpha_1 \rangle = 0.0050 \pm 0.0276$.
%After weighting, $\langle \alpha_1\rangle  = 0.0025 \pm 0.0205$. The $\alpha_1$ error bar again shrinks by 26\%, a substantial amount.
%

%We compare the distribution of the 1000 best-fit $\alpha_0$ and $\alpha_1$ values before and after using the $z$-weights. 
%Histograms of $\alpha_0$ and $\alpha_1$ from the ``unweighted" and ``weighted" fits to the 1000 mocks
%after reconstruction are shown in Fig. \ref{fig:recon_hist}. 
%The top panels are for the unweighted mocks. 
%The bottom panels are from simultaneously fitting the ``unweighted" and ``$x$-weighted" correlation functions. 
%We see that the best-fit $\alpha_1$ points come closer to the central value after weighting, giving us a better
%measurement of $\alpha_1$. 

\begin{figure*}
\centering 
\includegraphics[width=\textwidth]{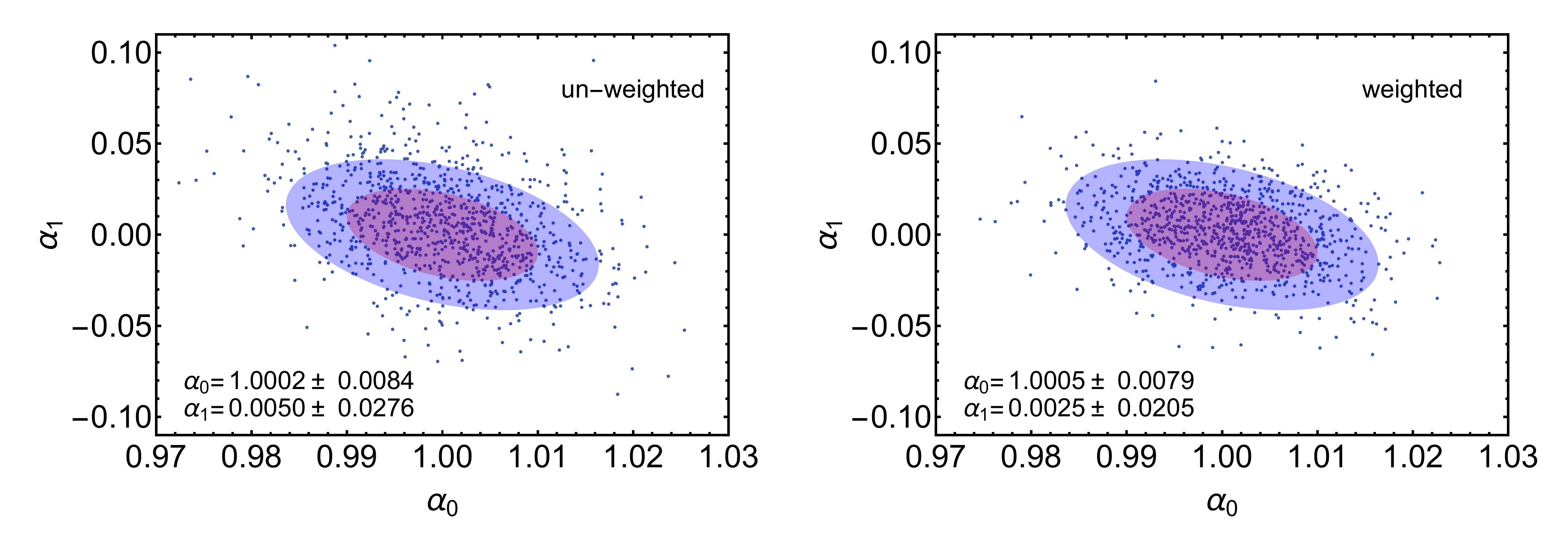}
\caption{The best-fit $\alpha_1$ versus $\alpha_0$ from the fits to 1000 individual mocks after reconstruction, 
assuming a pivot redshift $z_0 = 0.57$ in the analysis.
The left panel shows best-fit values from the ``unweighted" fits. The right panel is the same plot from ``weighted" fits. 
As expected, redshift weighting reduces the scatter of $\alpha_1$. 
The red and blue contours are $1 \sigma$ and $2 \sigma$ contours based on a Fisher forecast.}
\label{fig:scatter}
\end{figure*}

Fig.~\ref{fig:scatter} shows the scatter plot of $\alpha_0$ and $\alpha_1$ we obtain from 1000 mocks post-reconstruction. 
The left panel is from ``unweighted" fits and the right panel is after weights are being applied. 
We see that the two parameters are not highly correlated at this choice of the pivot redshift. 
We also plot the 1 and 2$\sigma$ error ellipse predicted from a Fisher matrix calculation (see Sec. \ref{sec:fisher} below)
 in both panels. 
The ellipses in the two panels are of the same size. One can see from the ``weighted" scatter plot 
that most of the best-fit $(\alpha_0, \alpha_1)$ points
fall within the 2$\sigma$ contour. This indicates that 
redshift weighting helps shrink the errors down towards the forecasted level. 

\begin{figure*}
\begin{minipage}{0.45\textwidth}
	\includegraphics[width=\textwidth]{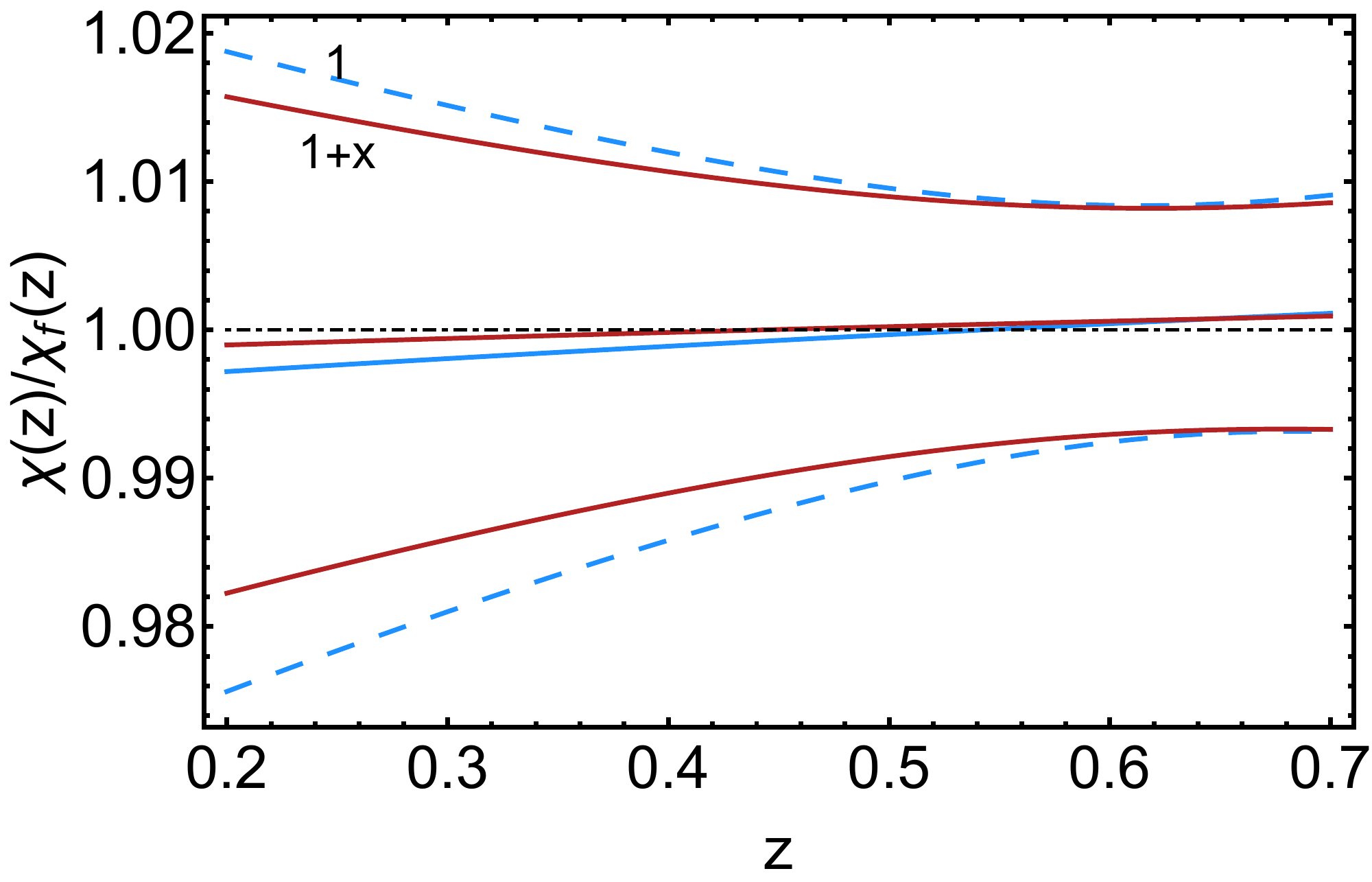}
	\end{minipage} 
	\hspace{0.05\textwidth}
	\begin{minipage}{0.45\textwidth}
	\includegraphics[width=\textwidth]{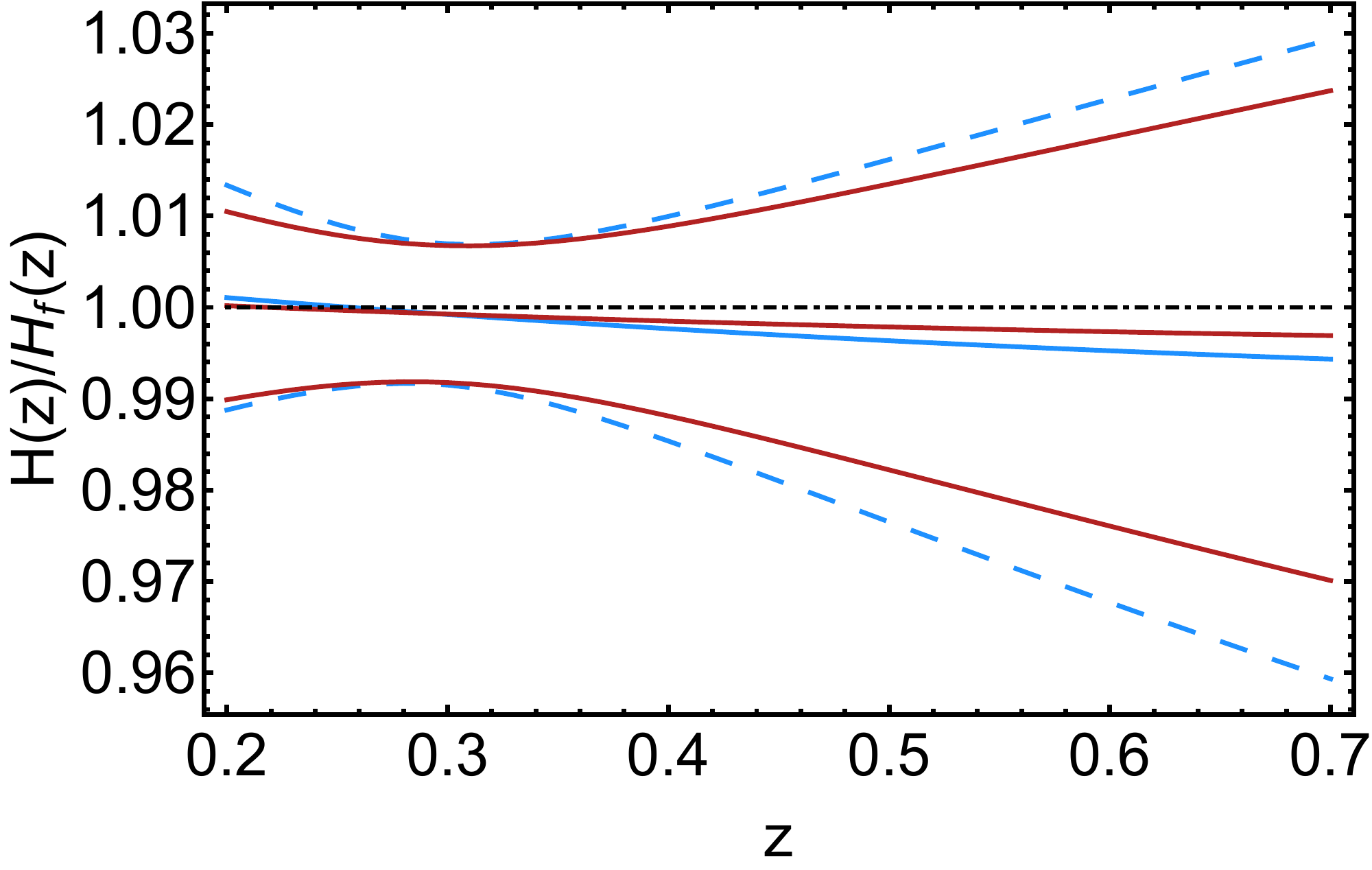}
	\end{minipage}
\caption{Distance $\chi(z)$ (left panel) and $H(z)$ measurements (right panel) from 1000 reconstructed mocks. 
Upper lines and bottom lines correspond to 1 standard deviation above and below the average (middle lines). 
The dashed [blue] line is from ``unweighted" fitting of the mocks and the solid [red] line is from
``weighted" fits where we simultaneously fit the unweighted and $x$-weighted correlation functions. 
The $z$-weights are effective in generating an unbiased and more accurate measurement of both the distance 
and Hubble parameter. 
}
\label{fig:recon_ratio}
\end{figure*}

With $\alpha_0$ and $\alpha_1$ in hand, we reconstruct the distance-redshift relation from Eq.~\ref{eq:chi_ratio}.
Similarly, we also reconstruct the Hubble parameter from Eq.~\ref{eq:h_ratio}. 
For each reconstructed mock, we use these best-fit $\alpha_0$ and $\alpha_1$ parameters to calculate the 
two relations and calculate the average and the scatter of each relation. We plot the reconstructed $\chi(z)/\chi_f(z)$
and $H(z)/H_f(z)$ with $1\sigma$ error in Fig. \ref{fig:recon_ratio}. The plots show the reconstructed relations from both
the ``unweighted" fits and the ``weighted" fits. Both $\chi(z)/\chi_f(z)$ and $H(z)/H_f(z)$ are centered around 1 at all redshifts,
suggesting applying the redshift weights give unbiased distance and Hubble parameter measurements. 
From the figures, we also find that weighting allows us to measure both
$\chi$ and $H$ to higher precision. The error of $\chi(z)/\chi_f(z)$ is smallest at higher redshifts. 
This reflects the fact that our sample is most concentrated 
at close to its ``effective redshift".

%Although we expect reconstruction to alleviate non-linearity, the measured $\chi(z)/\chi_f(z)$ still shows
%a slight offset from 1. This is not too concerning as we expect there is still some
%non-linear structure growth to cause small shift the BAO peak \citep{PadmanabhanWhite09, Mehta11}.

%\begin{figure}
%
%\caption{ $H(z)$ measurement from 1000 reconstructed mocks. Upper lines and bottom lines are 1 standard deviation. }
%\label{fig:recon_H}
%\end{figure}

\subsection{Robustness of Fits}

The fitting results above have assumed our default choices of fiducial cosmology, RSD streaming parameters, and galaxy bias. 
We explore the effects of varying these below. 

\subsubsection{Pivot Redshift}

We repeat the analysis by assuming a different pivot redshift $z_0 = 0.4$. The
weights are different from the set computed for $z_0 = 0.57$ since the weights are defined relative
to the comoving distance at the pivot redshift. 

\begin{figure*}
\centering 
\includegraphics[width=\textwidth]{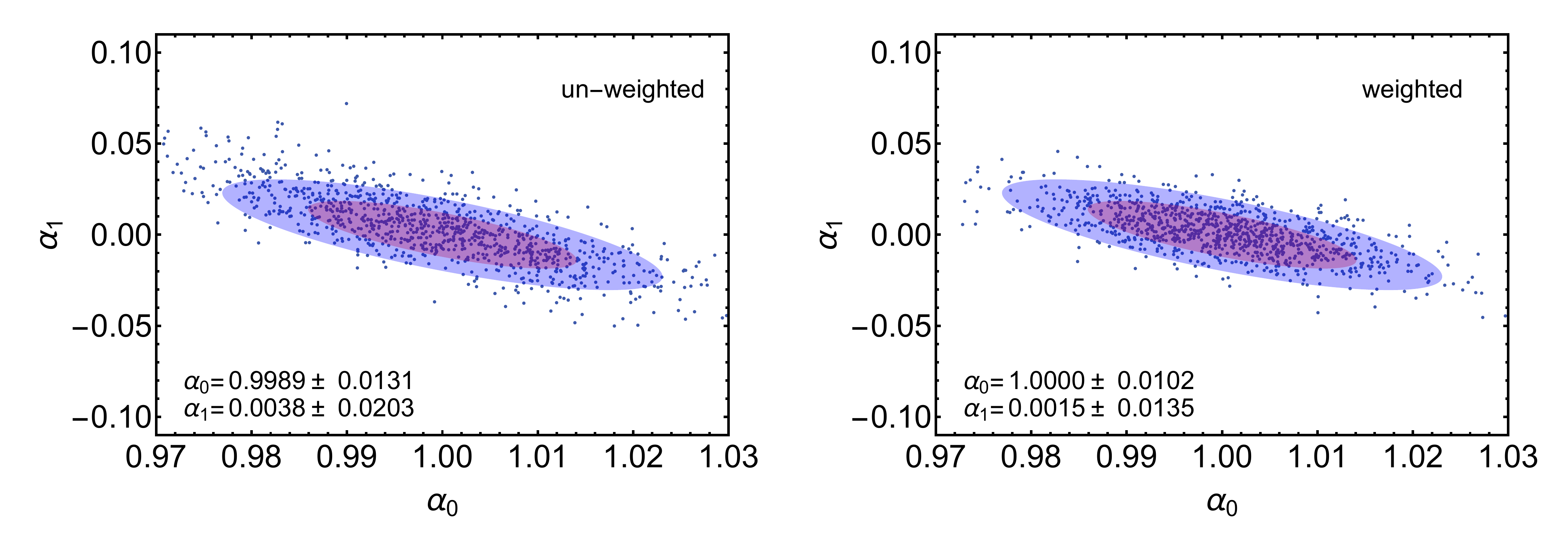}
\caption{The same plot as Fig. \ref{fig:scatter} while assuming the pivot redshift $z_0 = 0.4$ in the analysis. 
The scatter plot shows the same trend as in the $z_0 = 0.57$ case that redshift weighting makes the points come closer.}
\label{fig:scatter_z04}
\end{figure*}

We fit the 1000 reconstructed mocks assuming $z_0 = 0.4$ and summarize the statistics in the scatter plot in
Fig. \ref{fig:scatter_z04}. 
The measurements are still consistent with $\langle \alpha_0 \rangle = 1$ and $\langle \alpha_1 \rangle = 0$
within uncertainties. This confirms that weighting yields non-biased measurements of both parameters. 
In addition, redshift weighting again demonstrated efficiency in lowering the standard deviation of $\alpha_0$ and $\alpha_1$. 
The error on $\alpha_0$ is larger than the $z_0 = 0.57$ case while the error on $\alpha_1$ is smaller. 
Furthermore, the scatter plot shows clear correlation between the two parameters at this choice of pivot redshift.

%
%\begin{figure*}
%\centering
%\includegraphics[width=\textwidth]{plots/recon_plots/tot_histogram_no_b2_z04.pdf}
%\caption{  Histograms of best-fit $\alpha_0$ and $\alpha_1$ from the fits to the 1000 reconstructed mocks by assuming
%the pivot redshift $z_0 = 0.4$. 
%The top panel is for the ``unweighted" fits. The bottom panel is from ``weighted fits", 
%where we simultaneously fit the ``unweighted" and ``$x$-weighted" correlation functions. 
%The mean and standard deviation of $\alpha_0$ and $\alpha_1$ are labeled on each panel. 
%The $z$-weights are effective in bring down the error on $\alpha_1$. 
%The solid line is a Gaussian centered at the mean value of best-fit $\alpha$ values. 
%The width of the Gaussian is determined by a Fisher matrix forecast. 
%One sees that weighting brings the error down to close to the forecasted values, 
%suggesting its effectiveness
%in extracting nearly all the information from the mocks. 
%}
%\label{fig:recon_hist_z04}
%\end{figure*}

\begin{figure}
	\includegraphics[width=0.5\textwidth]{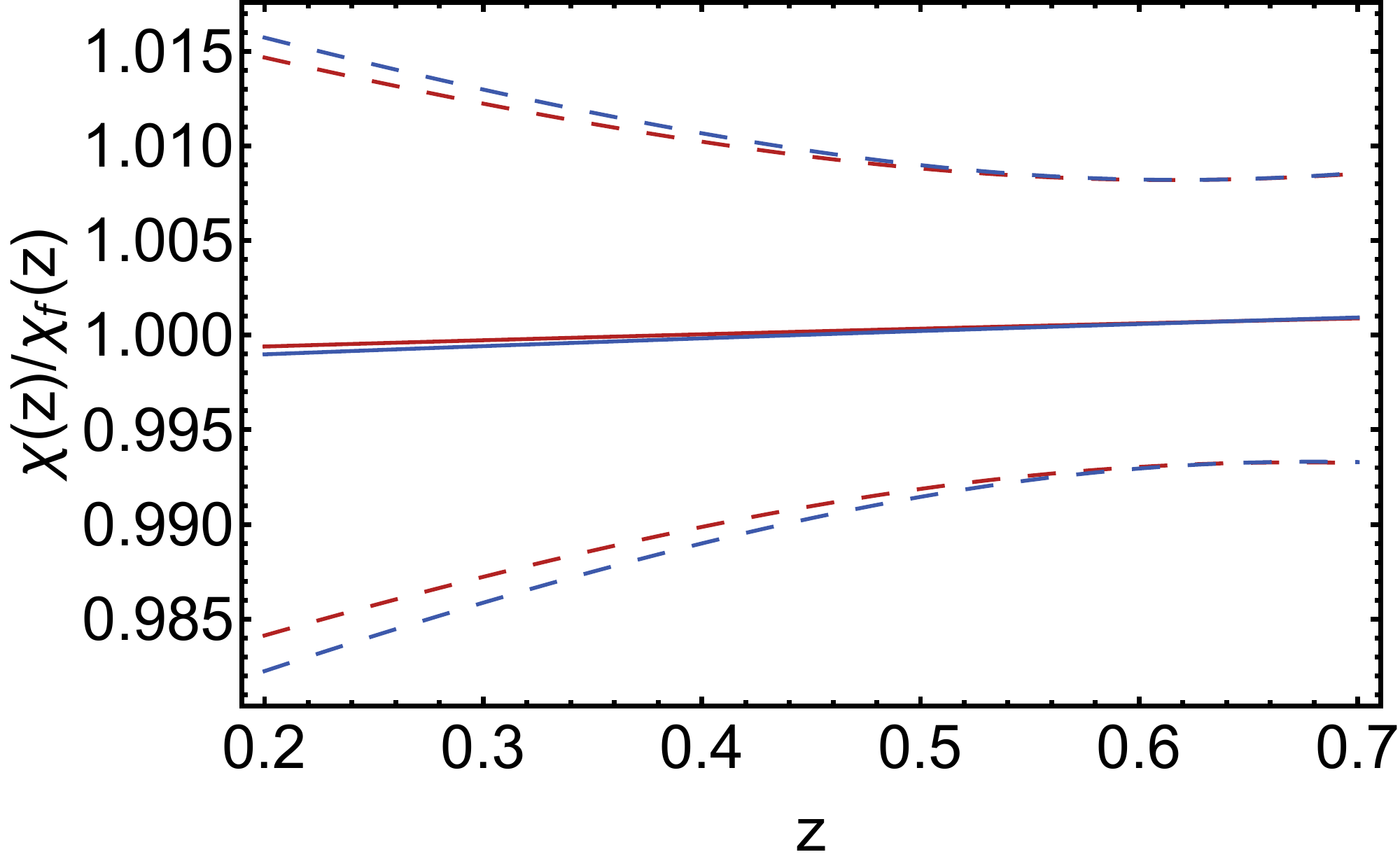}
	\includegraphics[width=0.5\textwidth]{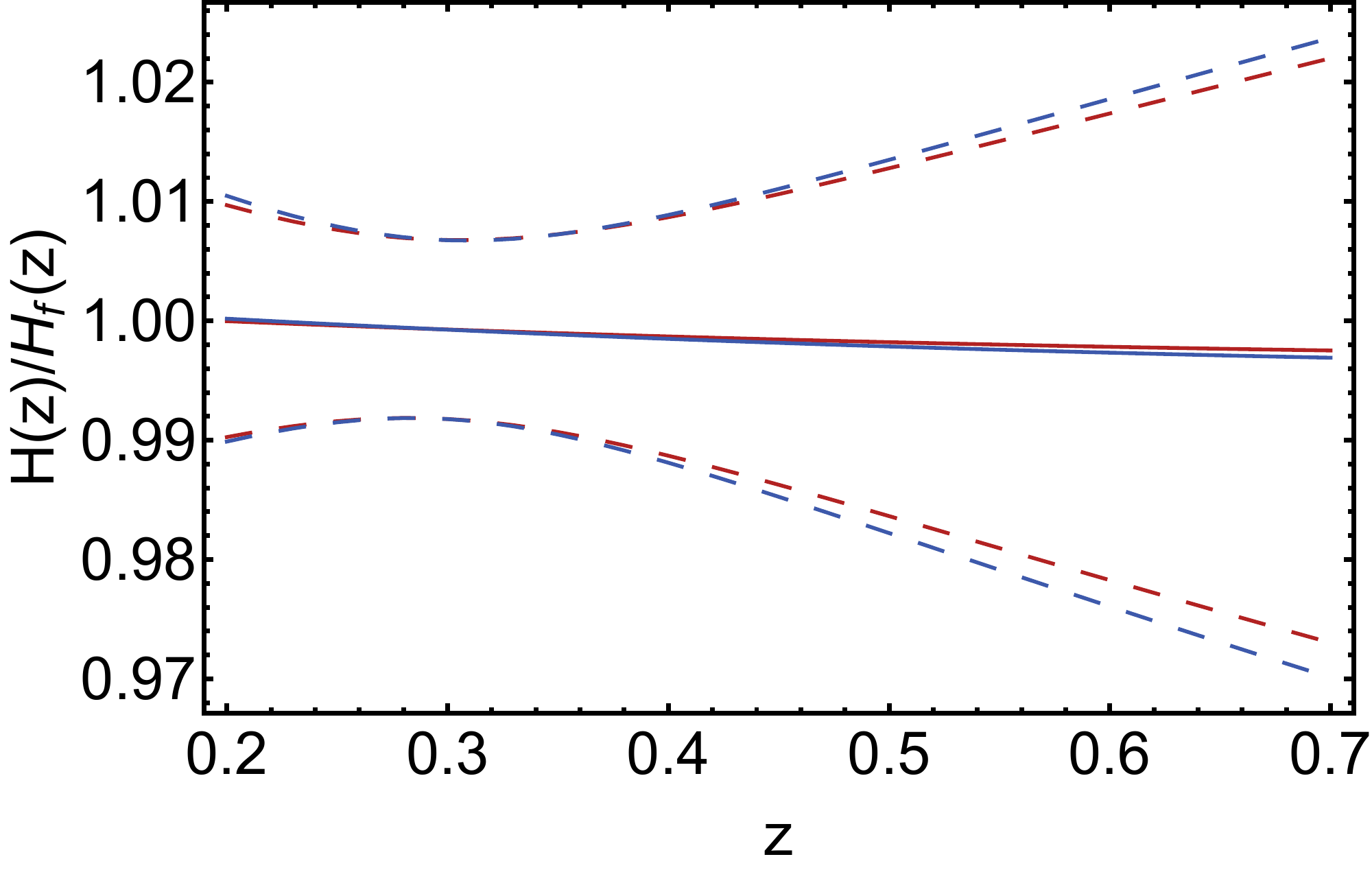}
\caption{Comparison of $\chi(z)$ (top panel) and $H(z)$ measurements (bottom panel) 
under two different pivot redshifts, $z_0 = 0.57$ (blue) and $z_0 = 0.4$ (red).
The fits to 1000 reconstructed mocks are done by using the ``weighted" fits. 
Upper lines and bottom lines correspond to 1 standard deviation above and below the average. 
We see that the reconstructed relations using two different pivot redshifts 
are almost identical, with the $z_0 = 0.4$ case doing slightly better. 
}
\label{fig:pivot_comparison}
\end{figure}

We reconstruct the distance-redshift relation and Hubble parameter based on the ``weighted" fits and compare them
against the $z_0 = 0.57$ results. The comparison is summarized by Fig. \ref{fig:pivot_comparison}. 
The analyses using two different pivot redshifts 
give almost identical reconstructed distance and Hubble parameter measurements. 

\subsubsection{Fiducial Cosmology}
\label{sec:fid_cosmo}
We test the robustness of the fitting routine and the gain in redshift weighting by using a 
fiducial cosmology that is different from the QPM cosmology. 
We pick a flat cosmology with $\Omega_m = 0.25$.
We fix $\Omega_m h^2 = 0.1421$ and $\Omega_b h^2 = 0.02247$ to be
the same as the QPM cosmology so that the sound horizon stays the same.  

Under this fiducial cosmology and pivot redshift $z_0 = 0.57$, we expect $\alpha_0 = 1.0599$
and $\alpha_1 = -0.0161$. 
Fitting the 1000 mocks yields $\alpha_0 = 1.0603 \pm 0.0084$ and $\alpha_1 = -0.0145 \pm 0.0203$, consistent with the 
expected values within uncertainties. This indicates the analysis and measurements are unbiased 
when the assumed fiducial cosmology differs from the true (simulation) cosmology. 

\subsubsection{Galaxy Bias Model} 

Our derivation of the redshift weights assumes a constant galaxy bias across redshifts. 
However, measuring the galaxy bias from small-scale clustering reveals a bias varying with redshift. 
The variation is rather mild, ranging from 1.65 to 1.8 in the redshift range $z = 0.2$ to $0.7$. 
This variation not only makes the default weights not optimal, 
it potentially can also bias the distance and Hubble measurement. 
We explicitly test for the effect by re-running the fits but assuming a constant galaxy bias $b(z) = 1.7$. 
The results (as presented in Table \ref{tab:bestfit_tab}) turn out to be almost identical to the default fits within uncertainty. 

\subsubsection{Including $\alpha_2$}
\label{sec:including_a2}

In the default fits, we have held $\alpha_2$ to be fixed at $0$.
%at its expected theory values according to Eq. \ref{eq:a2_theory}. 
However, the expected $a_2$ does not vanish when the fiducial cosmology differs from the true (QPM in our case) cosmology. 
The exclusion of $\alpha_2$ as a fitting parameter is equivalent to approximating the distance-redshift relation paramtrization to the first order.
%In the default fits, we have excluded $\alpha_2$ as a fitting parameter and held it fixed at $0$. This is equivalent to approximating distance-redshift relation parameterization to the first order. 
This approximation can potentially bias the measured $\alpha_0$ and $\alpha_1$, and in turn, bias the distance and Hubble parameter measurements. 
We explicitly test for such an effect by re-running the fits and including $\alpha_2$ as a fitting parameter. 
The fits assume a flat fiducial cosmology with 
$\Omega_m = 0.25$ (as in Sec. \ref{sec:fid_cosmo}). Under this cosmology, 
we expect $\alpha_0 = 1.0599$, $\alpha_1 = -0.0161$, and 
$\alpha_2 = 0.0018$. 
The fits yield $\alpha_0 = 1.0605\pm0.0083$, $\alpha_1 = -0.0155\pm0.0205$, and $\alpha_2 = -0.0175\pm0.1521$, all consistent with the expected theory values within uncertainty. 
%We have included the $x^2$-weighted correlation functions in the fits as $\alpha_2$ will be better constrained under this inclusion.  
The measured $15\%$ error on the $\alpha_2$ measurements suggests it 
cannot be well constrained by these data. 
Comparing the fitting results that assume $\alpha_2 = 0$ with 
our $\alpha_0$ and $\alpha_1$ measurements that includes $\alpha_2$ 
as a fitting parameter, we see that the former is unbiased within 
uncertainty. The reason is that the expected $\alpha_2$ is very close to 0. 
This is true for other reasonable fiducial cosmologies. 
In addition, we reconstruct the distance-redshift relation and Hubble parameter 
with the full quadratic expansion in Eq. \ref{eq:chi_ratio} and Eq. \ref{eq:h_ratio} and find the results are 
almost identical to assuming $\alpha_2 = 0$. 
Hence we claim in general the default fits with $\alpha_2$ forced to be zero are sufficient 
and unbiased within uncertainty.

\subsection{Comparison with Fisher Matrix Forecasts}
\label{sec:fisher}

The Fisher matrix is a commonly used tool in estimating errors from a planned survey. 
Inverting the Fisher matrix gives the parameter covariance matrix. It serves as a marker 
for the theoretical lower limit of errors measured from a planned survey. We describe the 
details that go into a Fisher matrix calculation and compare the errors from our ``weighted" fits 
to the Fisher matrix forecasts. 

We break the redshift range of the survey [0.2, 0.7] into 50 bins, each with width $\Delta z = 0.01$. 
The volume of each slice is computed according to 
\begin{equation}
	\Delta V_z = \frac{\chi_f^2(z)}{H_f(z)}\Delta z \Delta \Omega
\end{equation}
where $\Delta\Omega$ is the angle covered by the BOSS DR12 area. 

In each redshift slice, we calculate the Fisher matrix for $\chi(z)$ and $1/H(z)$ according to \cite{Seo+Eisenstein}. 
We assume 
$\beta = 0$, $\Sigma_s = 0$, and $\Sigma_\perp = \Sigma_\parallel = 4.3 h^{-1}{\rm Mpc}$ 
post-reconstruction motivated by fits to the average correlation function. 

We then rotate the basis into $\alpha_0$, $\alpha_1$, and $\alpha_2$ through a linear transformation :
\begin{equation}
	{\cal F}_{\alpha_0,\alpha_1,\alpha_2} = J^T {\cal F}_{\chi,H^{-1}} J
\end{equation}
where $J$ is the Jacobian matrix 
\begin{equation}
	J = 	\left( 
	\begin{array}{ccc}
		\frac{\partial \chi}{\partial \alpha_0} & \frac{\partial \chi}{\partial \alpha_1} & \frac{\partial \chi}{\partial \alpha_2} \\
		\frac{\partial H^{-1}}{\partial \alpha_0} & \frac{\partial H^{-1}}{\partial \alpha_1} & \frac{\partial H^{-1}}{\partial \alpha_2} 
	\end{array} \right)
\end{equation}
If one focuses on $\alpha_0$ and $\alpha_1$ and have $\alpha_2$ fixed to be 0, the Jacobian matrix is made up of the first two columns. 

Using Eq. \ref{eq:chi_ratio} and Eq. \ref{eq:h_ratio},  we compute the Jacobian matrix as
\begin{equation}
	J
	=
	\left( 
	\begin{array}{ccc}
		\chi_f(z)& \chi_f(z) x & \frac{1}{2}\chi_f(z) x^2\\
		\frac{1}{H_f(z)} & \frac{1+2x}{H_f(z)} &  \frac{x+\frac{3}{2}x^2}{H_f(z)}
	\end{array} \right).
\end{equation} 
Once we have calculated the Fisher matrix for $\alpha_0$, $\alpha_1$, and $\alpha_2$
in each redshift slice, we combine the errors calculated in these slices through inverse variance weighting. 
This corresponds to a sum of the Fisher matrices 
\begin{equation}
	F = \sum_z F(z).
\end{equation}
Inverting the total Fisher matrix gives the parameter covariance matrix $C = F^{-1}$. 

Focusing on the two parameter $(\alpha_0, \alpha_1)$ case, 
the Fisher matrix calculation for $z_0 = 0.57$ yields the estimated errors of $\alpha_0$, $\alpha_1$ 
to be $0.66\%$ and $1.67\%$ respectively. 
For the $z_0 = 0.4$ case, the Fisher forecast yields $0.93\%$ error on $\alpha_0$ and $1.22\%$ on $\alpha_1$.
These errors are about $10\%$ to $20\%$ lower than what we have measured 
from the weighted fits. In the three parameter $(\alpha_0, \alpha_1, \alpha_2)$ case, the errors of $\alpha_0$ and $\alpha_1$ remain
comparable as in the two parameter case. The estimated error of $\alpha_2$ is $11\%$, suggesting $\alpha_2$ cannot be
well constrained by these data. 

%\begin{figure}
%	\includegraphics[width=0.5\textwidth]{plots/recon_plots/a_error2.pdf}
%\caption{Projected errors of $\alpha_0$ and $\alpha_1$ from a Fisher forecast as we vary
%the pivot redshift. A higher pivot redshift allows a better measurement of $\alpha_0$ but a worse measurement 
%of $\alpha_1$. 
%}
%\label{fig:a0a1_fisher_zpivot}
%\end{figure}

To analyze the impact from different choices of pivot redshifts, we calculate the errors on $\alpha_0$ and $\alpha_1$
for different pivot redshifts. We find that
a higher pivot redshift allows a better measurement of $\alpha_0$ but a worse $\alpha_1$.  
We also find that the correlations between the two parameters 
$\rho_{\alpha_0 \alpha_1} = C_{\alpha_0 \alpha_1} / \sqrt{C_{\alpha_0 \alpha_0} C_{\alpha_1 \alpha_1}}$
increases from $\rho_{\alpha_0 \alpha_1} = -0.9$ at $z=0.2$ to $\rho_{\alpha_0 \alpha_1} = 0.1$ at $z=0.7$. 
They decorrelate at redshift $z=0.68$. We calculate the forecasted errors of $\chi(z)/\chi_f(z)$ and 
$H(z)/H_f (z)$ at different pivot redshifts and found
them to be insensitive to the choice of the pivot redshift. 
The error of $\chi(z)/\chi_f(z)$ reaches as low as $0.61\%$ at around $z = 0.68$. 
The error of $H(z)/H_f(z)$ is smallest at roughly $z=0.3$. 
These are all consistent with the mock results within $10\%$ to $20\%$. 

The Fisher matrix calculation also allows us to gain insight into the constraining power of
 $D_A$ and $H$ measurements on $\alpha_0$ and $\alpha_1$. 
We make the following experiment in our Fisher matrix calculation. 
In each redshift slice, we increase the error of $H$ while keeping the error
of $\chi$ the same. 
Table \ref{tab:fisher_exp} lists the estimated $\alpha_0$ and $\alpha_1$ errors with the $H$ errors increased
by 2 fold, 10 fold, and 1000 fold in each redshift slice. When we increase the error of the Hubble parameter $H$ by 2,
we find that $\alpha_0$ error goes up by $10\%$ while the $\alpha_1$ error quickly worsens. This suggests the $H$ 
measurement is important for constraining $\alpha_1$ to high precision.  
As we continue to increase $H$ errors, $\alpha_0$ and $\alpha_1$ errors continue to grow. 
The case where the $H$ error is increased by a factor of 1000 mimics the case in which the survey only affords $D_A$ measurements but not $H$. In this case the information is predominantly from $D_A$ measurements. The estimated error of $\alpha_0$ is at the $1\%$ level and $\alpha_1$ error is $4.5\%$.

\begin{table}
%\centering
\begin{tabular}{| c | c | c | }
    \hline
     $H$ error increased by &  $\alpha_0$ error (in $\%$) & $\alpha_1$ error (in $\%$) \\ \hline
     Original &    $0.66$ & $1.6$  \\ \
     2x & $0.72$ & $2.5$ \\ \
     10x & $0.94$ & $4.1$ \\ \
     1000x & $1.04$ & $4.5$ \\ 
   \hline
\end{tabular}
\caption{Variation of the estimated $\alpha_0$ and $\alpha_1$ errors with Hubble parameter errors
increased by 2 folds, 10 folds, and 1000 folds.}
\label{tab:fisher_exp}
\end{table}

\section{Discussion}

%We apply the redshift weighting technique in \cite{Zhu15} to the clustering of galaxies 
%from 1000 QuickPM (QPM) mock simulations post-reconstruction and achieve a $0.75\%$
%measurement of the angular diameter distance $D_{A}$ at $z=0.64$ 
%and the same precision for Hubble parameter $H$ 
%at $z=0.29$. These QPM mock catalogs are designed to mimic the clustering and noise level of the 
%Baryon Oscillation Spectroscopic Survey (BOSS) Data Release 12 (DR12). 
%We implement the redshift weights proposed in \cite{Zhu15} to compress the correlation functions 
%in the redshift direction onto a set of weighted correlation functions. These estimators give unbiased $D_{A}$ and $H$ 
%measurements at all redshifts within the range of the combined sample. 
%We demonstrate the effectiveness of redshift weighting in improving the distance and Hubble parameter estimates. 
%Instead of measuring at a single `effective' redshift as in traditional analyses, we report our $D_{A}$ and $H$ measurements at all redshifts. 
%The measured fractional error of $D_{A}$ ranges from $1.53\%$ at $z=0.2$ to $0.75\%$ at $z=0.64$. 
%The fractional error of $H$ ranges from $0.75\%$ at $z=0.29$ to $2.45\%$ at $z=0.7$. 
%Our measurements are consistent with a Fisher forecast to within $10\%$ to $20\%$ depending on the pivot redshift. 
%We further show the results are robust against the choice of fiducial cosmologies, galaxy bias models, and RSD streaming parameters. 

This paper presents the results of applying redshift weighting as proposed in \cite{Zhu15} to BAO analyses. 
Different from previous BAO analyses, redshift weighting
allows us to analyze a full sample without the need of splitting the sample into multiple redshift bins. 
We validate the method on a set of 1000 QPM mocks 
tailored to mimic the clustering noise level of BOSS DR 12. 

We approximate the distance-redshift relation, relative to a fiducial model, by a quadratic function. 
By measuring the coefficients from the mocks, 
we then reconstruct the distance and Hubble parameter measurements from the expansion. 
Our approach thus gives measurements of $D_A(z)$ and $H(z)$ at all redshifts within the range of the sample. 
This is different from previous analyses in which only measurements at the ``effective redshift" are given. 
Our fits assume the Hubble parameter to be the inverse derivative of the comoving distance. 
%We derived a complementary parameterization of the Hubble parameter 
%by taking the inverse derivative of our parameterized distance-redshift relation.
We are thus jointly measuring $D_A$ and $H$ with this
additional constraint in place. 
This differs from traditional analyses in which $D_A$ and $H$ are measured separately.

The key advantage of redshift weighting is the optimized use of the full sample. 
We compress the information in the redshift direction 
into a small number of `weighted correlation functions'. 
These weighted estimators preserve nearly all the BAO information 
without diluting the signal-to-noise per measurement. 
We found that fitting these weighted estimators improves the distance and Hubble parameter measurements. 
Our mock results yield a $0.75\%$ $D_A$ measurement at $z=0.64$ and the 
same precision for $H$ at $z=0.29$. We can compare our results to the results of \cite{Cuesta16} who
analyzed a similar sample by splitting into 2 redshift bins. In that work, they measured $D_A$ and $H$ 
with $2.5\%$ and $5.2\%$ uncertainty respectively for the LOWZ sample ($0.2 < z < 0.43$), and $1.6\%$ and $3.1\%$ for CMASS ($0.43 < z < 0.7$). 

We demonstrate that our method is unbiased
and robust against the choices of fiducial cosmologies, pivot redshift, RSD streaming 
parameters, and galaxy bias models. 
We have also extended the fits to include the second order term in the expansion of our distance-redshift parametrization 
and found the results to be almost identical. We thus claim the default fits with the first order of the parametrization is sufficient. 

%
%This could be due to that isotropic reconstruction convention generates a BAO signal along the line of sight that 
%cannot be modeled by a simple Gaussian damping model \citep{Seo2015}. 
%Although the modified model gives a much better fit to the average correlation function,
% we don't see an improvement in the distance and Hubble measurements from individual mocks.
%Another possible source of discrepancy may be due to the Fisher forecast being overly optimistic when 
%we have assumed the redshift shells to be independent.
%The improvement on distance constraint is most significant at lower redshifts, while being most prominent 
%for Hubble parameter at higher redshifts.  
%\cite{Cuesta16} performed BAO analyses of the same mock catalogs using the traditional methods 
%and reported 

We compare our results with a Fisher matrix forecast. 
Our results are $10\%$ worse than the estimated Fisher errors. 
We experiment with the Fisher matrix calculation by degrading $H$ measurements by 1000 fold in each
redshift slice and re-estimate $\alpha_0$ and $\alpha_1$ uncertainties.  
At pivot redshift $z=0.57$, the $\alpha_0$ and $\alpha_1$ errors degrade from
$0.66\%$ and $1.6\%$ to $1\%$ and $4.5\%$. 
This exercise allows us to estimate how much information $D_A$ measurement alone 
affords in constraining $\alpha_0$ and $\alpha_1$. This estimate is potentially useful for photometric surveys. 

Our algorithm and results have important implications for BAO measurements from current and
future redshift surveys. 
The same technique has also been proposed for analyzing the 
RSD signal and the combined BAO and RSD signal \citep{Ruggeri2016, Zhao2016}. 
As future surveys will probe large volumes, 
covering wide ranges in redshift, we expect redshift weighting to be very useful. 
We plan on continuing to develop this
approach in future work by applying it to existing surveys. 

\label{sec:discussion}

\section{Acknowledgments}

We would like to thank Will Percival for helpful conversations. 
This work was supported in part by the National Science Foundation under 
Grant No. PHYS-1066293 and the hospitality of the Aspen Center for Physics.
NP and FZ are supported in part by a DOE Early Career Grant DE-SC0008080. 

Funding for SDSS-III has been provided by the Alfred P. Sloan Foundation, the Participating Institutions, the National Science Foundation, and the U.S. Department of Energy Office of Science. The SDSS-III web site is http://www.sdss3.org/.

SDSS-III is managed by the Astrophysical Research Consortium for the Participating Institutions of the SDSS-III Collaboration including the University of Arizona, the Brazilian Participation Group, Brookhaven National Laboratory, Carnegie Mellon University, University of Florida, the French Participation Group, the German Participation Group, Harvard University, the Instituto de Astrofisica de Canarias, the Michigan State/Notre Dame/JINA Participation Group, Johns Hopkins University, Lawrence Berkeley National Laboratory, Max Planck Institute for Astrophysics, Max Planck Institute for Extraterrestrial Physics, New Mexico State University, New York University, Ohio State University, Pennsylvania State University, University of Portsmouth, Princeton University, the Spanish Participation Group, University of Tokyo, University of Utah, Vanderbilt University, University of Virginia, University of Washington, and Yale University.

Some of the codes in this paper made use of the Chapel programming language\footnote{\texttt{http://chapel.cray.com}}.

\bibliographystyle{mn2elong}
\bibliography{paper2}

\appendix

\section{Expected value of $\langle \chi^2 \rangle $}

Consider the $p$-dimensional observations, $\bf{x}$. 
Each entry of ${\bf x}$ is a random variable with a gaussian distribution 
with mean $0$ and standard deviation $1$, denoted as 
$x \sim {\cal N}(0, 1)$. 

We compute the sample covariance matrix from $d$ independent samples, ${\bf x}_i$ where $1\le i \le d$. 
\begin{equation}
	{\bf C} = \frac{1}{d-1} \sum_{i=1}^d ({\bf x}_i) ({\bf x}_i)^t
\label{eq:cov}
\end{equation}
where the superscript $t$ is the transpose. An unbiased estimate of the precision matrix is then 
\begin{equation}
	\hat{{\bf \Psi}} = \left(1-\frac{p+1}{d-1}\right){\bf C}^{-1}.
\label{eq:precision}
\end{equation}
Note that both the covariance matrix ${\bf C}$ and precision matrix ${\bf \Psi}$ are $p\times p$ matrices. 

The average $\chi^2$ is given by
\begin{equation}
	\langle \chi^2 \rangle = \frac{1}{d} \sum_{i=1}^d {\rm tr} \left({\bf x}_i^t \hat{\bf \Psi} {\bf x}_i \right)
	= \frac{1}{d} \sum_{i=1}^d {\rm tr} \left({\bf x}_i {\bf x}_i^t \hat{\bf \Psi} \right).
\label{eq:average_chi2}
\end{equation}
In the second equality, we have used the cyclic property of trace. 
Inserting Eq. \ref{eq:cov} and \ref{eq:precision} into Eq. \ref{eq:average_chi2}, 
we obtain the expected average $\chi^2$ as
\begin{align}
	\langle \chi^2 \rangle & = \frac{1}{d} {\rm tr}\left((d-1){\bf C}\hat{\bf \Psi}\right) \\
	& =  \frac{d-1}{d} \left(1-\frac{p+1}{d-1}\right)p. 
	\label{eq:expected_chi2}
\end{align}
The above calculation can be generalized for other distributions. The key message remains the same - that 
if we fit $d$ independent samples by using the covariance matrix calculated from the same samples, 
the expected $\langle \chi^2 \rangle$ and the degree-of-freedom $p$ are related by Eq. \ref{eq:expected_chi2}.

\end{document}